\documentclass{ws-rv975x65}
\usepackage{subfigure}     
\usepackage{ws-rv-van}     
\makeindex
\renewcommand{\vec}[1]{\mbox{\boldmath $#1$}}

\begin{document}

\begin{flushleft}

This chapter contains the following copyrighted materials: 

\begin{picture}(400,10)
\put(0,0){\line(1,0){360}}
\end{picture}

\end{flushleft}

\noindent
Reprinted Fig. 3 with permission from 
T. Nakamura et al., Phys. Rev. Lett. 96, 252502 (2006).  
Copyright 2006 by the American Physical Society. 

\medskip

\noindent
Reprinted Fig. 4 with permission from 
T. Aumann et al., Phys. Rev.C59, 1252 (1999). 
Copyright 1999 by the American Physical Society. 

\medskip

\noindent
Reprinted Fig. 1 with permission from 
K. Miernik et al., Phys. Rev. Lett. 99, 192501 (2007). 
Copyright 2007 by the American Physical Society. 

\medskip

\noindent
Reprinted Fig. 2 with permission from 
A. Spyrou et al., Phys. Rev. Lett. 108, 102501 (2012). 
Copyright 2012 by the American Physical Society. 

\medskip

\noindent
Reprinted Fig. 3 with permission from 
I. Tanihata et al., Phys. Rev. Lett. 100, 192502 (2008). 
Copyright 2008 by the American Physical Society. 

\medskip

\noindent
Reprinted Fig . 1 with permission from 
A. Ozawa et al., Phys. Rev. Lett. 84, 5493 (2000). 
Copyright 2000 by the American Physical Society. 

\medskip

\noindent
Reprinted Fig . 2 with permission from 
I. Hamamoto, Phys. Rev. C69, 041306(R) (2004). 
Copyright 2004 by the American Physical Society. 

\medskip

\noindent
Reprinted Fig . 2 with permission from 
P. Adrich et al., Phys. Rev. Lett. 95, 132501 (2005). 
Copyright 2005 by the American Physical Society. 

\medskip

\noindent
Readers may view, browse, and/or download
material for temporary copying purposes only, provided these 
uses are for noncommercial personal purposes.
Except as provided by law, this material may not be 
further reproduced, distributed, transmitted, modified, adapted,
performed, displayed, published, or sold in whole or part, 
without prior written permission from the American
Physical Society.
\begin{flushleft}
\begin{picture}(400,0)
\put(0,0){\line(1,0){360}}
\end{picture}
\end{flushleft}

\noindent
Reprinted Fig. 2 from 
Phys. Lett. B268, M. Fukuda et al., 
Neutron halo in $^{11}$Be studied via reaction cross sections, 
p. 339 (1991), with permission from Elsevier. 



\medskip

\noindent
Reprinted Fig. 3 from 
Nucl. Phys. A693, A. Ozawa, T. Suzuki, and I. Tanihata, 
Nuclear size and related topics, 
p. 32 (2001), 
with permission from Elsevier. 

\medskip

\noindent
Reprinted Fig. 2 from 
Phys. Lett. B707, 
M. Takechi et al., 
Interaction cross sections for Ne isotopes towards the island 
of inversion and halo structures of $^{29}$Ne and $^{31}$Ne, 
p. 357 (2012), 
with permission from Elsevier. 

\medskip

\noindent
Reprinted Fig. 3 from 
Phys. Rep. 264, 
T. Otsuka and N. Fukunishi, 
Nuclear mean field on and near the drip lines, 
p. 297 (1996), 
with permission from Elsevier. 
\begin{flushleft}
\begin{picture}(400,0)
\put(0,0){\line(1,0){360}}
\end{picture}
\end{flushleft}


\chapter[Exotic nuclei far from the stability line]{Exotic 
nuclei far from the stability line}

\author[K. Hagino, I. Tanihata, and H. Sagawa]
{K. Hagino$^1$, I. Tanihata$^{2,3}$, and H. Sagawa$^4$}
\address{$^1$Department of Physics, Tohoku University, Sendai 980-8578, Japan \\
$^2$Research Center for Nuclear Physics (RCNP), Osaka University, 
Mihogaoka, Ibaraki, Osaka 5670047, Japan \\
$^3$
Research Center of Nuclear Science and Technology(RCNST) and School 
of Physics and Nuclear Energy Engineering, Beihang University,
No. 37 Xueyuan Road, Haidian District, Beijing, P. R. China 100191 \\
$^4$Center for Mathematics and Physics,  University of Aizu,
Aizu-Wakamatsu, Fukushima 965-8560,  Japan
}

\begin{abstract}
The recent availability of radioactive beams has opened up 
a new era in nuclear physics. 
The interactions and structure of exotic nuclei 
close to the drip lines have been studied extensively world wide, 
and it has been revealed that 
unstable nuclei, having weakly bound nucleons, 
exhibit characteristic features
such as a halo structure and a soft dipole excitation.
We here review the developments of the physics of unstable nuclei 
in the past few decades. 
The topics discussed in this Chapter include the halo and skin structures, the 
Coulomb breakup, the dineutron correlation, the pair transfer reactions, 
the two-nucleon radioactivity, the appearance of new magic numbers, 
and the pygmy dipole resonances. 
\end{abstract}
\body

\section{Introduction}

Until the middle of 1980's, nuclear physics 
had been developed by investigating primarily stable nuclei 
which exist in nature. Many facets of atomic nuclei had been revealed, 
which include a mass, density distribution, radius, shell structure, 
collective excitations, 
and various decay modes\cite{BB36,BM69,RS80}. 
As a natural question, however, it had been discussed already 
in the late 1960's a question on how many neutrons 
can be bound for a given number of proton\cite{G72,K72,V74}. 
The first international symposium on 
nuclei far from the stability line was held in 1966 at Lysekil, 
Sweden\cite{Lysekil66}, 
followed by the succeeding conference held in 1970 at CERN\cite{CERN70}. 
The questions which attracted nuclear physicists at that time include 
i) where is the neutron drip line? ii) do the nuclear models which 
were successful for stable nuclei work also for neutron-rich nuclei? 
and iii) the relevance to the r-process nucleosynthesis\cite{G72}. 
It is remarkable that already in 1966 the $^8$He nucleus was predicted to be 
stable by about 10 MeV against a dissociation to $^6$He + 2n\cite{GK66}. The 
$^{11}$Li nucleus was also predicted to be slightly unbound, that is, only 
by 0.6 MeV, with respect to the $^9$Li+2n configuration \cite{GK66}. 

\begin{figure}[t]
\centerline{\psfig{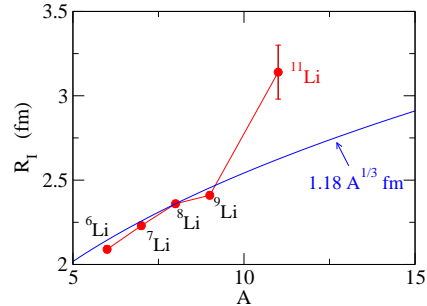}}
\caption{The matter radii of Li isotopes deduced from the measured 
interaction cross sections with a carbon target at 790 MeV/nucleon. 
The solid line shows the systematics known for stable nuclei, 
$R\propto A^{1/3}$, where $A$ is the mass number. 
The experimental data are taken from Ref. \citen{T85}.}
\end{figure}

The real start of the field of neutron-rich nuclei, however, was 
much later, that is, only in 1985, when  an anomalously large matter 
radius of the $^{11}$Li nucleus was experimentally discovered through the measurement 
of the interaction cross section \cite{T85}. 
The matter radius was found to deviate largely from the known systematics in stable 
nuclei, which scales as $A^{1/3}$ 
as a function of mass number $A$ of a nucleus (see Fig. 1). 
Together with a fact that the two-neutron separation energy, $S_{2n}$,  
is extremely small for this nucleus ($S_{2n}$ was known to be 190$\pm$ 
110 keV at the time of 1985 \cite{WAB85}, which has been updated to 
378$\pm$5 keV\cite{BAG08}), the large matter radius has been 
interpreted to be due to a long tail of the wave function for the 
weakly bound valence nucleons.\cite{HJ87}. 
This structure is referred to as {\it halo}, in which the density 
distribution of valence 
neutron(s) largely extends over the core nucleus. 
Since the proton and neutron density distributions are almost the same 
in stable nuclei, the discovery of the halo structure was a big surprise. 
Subsequently, the interpretation of the halo structure was 
supported also by the observed narrow momentum distribution 
of the $^9$Li nucleus due to the breakup of $^{11}$Li \cite{KYO88}. 
It has been recognized by now that 
this exotic structure is an important characteristic 
feature of neutron-rich nuclei, and it has attracted much attention. 

The physics of neutron-rich nuclei has now been one of the 
main current subjects of nuclear physics. 
In fact, new generation RI beam facilities (such as RIBF at RIKEN in Japan\cite{M10}, 
FAIR at GSI in Germany\cite{FAIR}, 
SPIRAL2 at GANIL in France\cite{SPIRAL2}, and FRIB at MSU in 
the USA\cite{FRIB}) have been, or will soon be, in operation in the world 
wide. 
In this chapter, we 
summarize the developments of the physics of unstable nuclei 
in the past few decades. In Sec. II, we first discuss properties of one-neutron 
halo nuclei, that is, nuclei with a tightly bound core nucleus 
and a loosely bound valence neutron. 
We particularly discuss the role of single-particle angular momentum and 
the Coulomb dissociation. 
In Sec. III, we treat two-neutron halo nuclei. The main focus will be put on 
the correlation between the valence neutrons. As possible probes for 
the correlation, we discuss the Coulomb breakup, the two-nucleon 
radioactivity, and the two-neutron transfer reactions. 
In Sec. IV, we consider heavier neutron-rich nuclei. The topics to be 
discussed include the matter radii and the neutron skin thickness, 
the odd-even staggering of interaction cross sections, 
alpha clustering, the shell evolution and the deformation, 
and the collective excitations such as the pygmy dipole resonances. 
Finally, we summarize the chapter in Sec. V. 

As the physics of neutron-rich nuclei is diverse, it is almost impossible 
to cover all the topics in this chapter. 
We would like the readers to refer also to 
review articles
\cite{MS93,T95,HJJ95,T96,T98,HT03,J04,JRFG04,IMKT10,NK12,AN03,Yahiro12,
CGDH06,PVKC07,PKGR12}, 
and references therein. 
 
\section{One-neutron halo nuclei}

\subsection{Role of single-particle angular momentum}

\begin{figure}[htb]
\centerline{\psfig{file=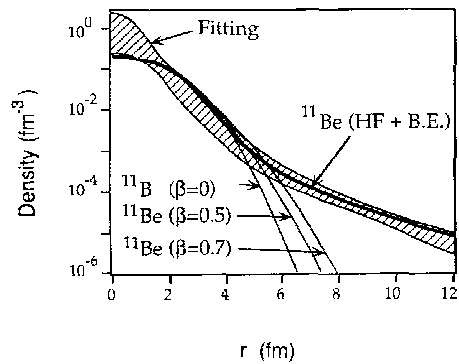,width=7cm}}
\caption{The density distribution of the $^{11}$Be nucleus 
which is consistent with the measured interaction cross sections (see 
the hatched area). Taken from Ref. \citen{Fukuda91}.}
\end{figure}

We first discuss properties of one-neutron halo nuclei, for which a 
weakly-bound valence 
neutron moves around a core nucleus. A typical example is $^{11}$Be, for which 
the interaction cross section has been found considerably 
large \cite{T85,T88},  similar to the $^{11}$Li nucleus. 
The one neutron separation energy, $S_n$, is as small as 
504$\pm$ 6 keV \cite{AW93}, which can be compared to the one neutron separation 
energy of {\it e.g.,} $^{13}$C nucleus, $S_n$=4.95 MeV. 
This suggests that the $^{11}$Be nucleus takes a halo structure. In fact, 
the large interaction cross section is consistent with a density 
distribution with a long tail, as has been shown in Ref. \citen{Fukuda91} (see 
Fig. 2). 

\begin{figure}[t]
\centerline{\psfig{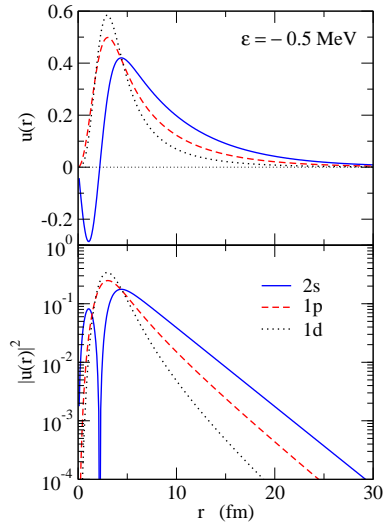}}
\caption{The radial wave functions $u(r)$ for the relative motion between 
the valence neutron and the core nucleus $^{10}$Be in the $^{11}$Be 
nucleus. 
The upper panel shows the wave functions on the linear scale, while the 
lower panel shows the square of the wave functions on the logarithmic 
scale. 
The solid, the dashed, and the dotted lines correspond to 
the wave functions for the 2$s$, 1$p$, and 1$d$ states, respectively. 
A Woods-Saxon shape is assumed for the mean-field potential, 
whose depth is adjusted for each angular momentum so that the single-particle 
energy is
$\epsilon=-0.5$ MeV }
\end{figure}

\begin{figure}[hbt]
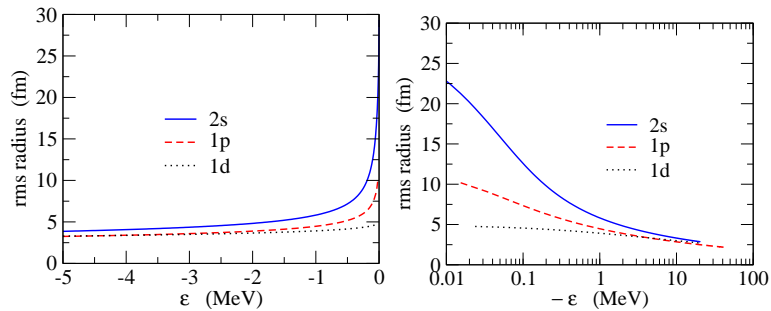

\psfig{file=fig4a.eps,clip,width=5cm}
\psfig{file=fig4b.eps,clip,width=5cm}
\caption{The root-mean-square (rms) radii for the 
2$s$, 1$p$, and 1$d$ states in the $^{11}$Be nucleus as a function 
of the single-particle energy $\epsilon$. The meaning of each line 
is the same as in Fig. 3. The right panel is the same as the 
left panel, but plotted in the logarithmic scale for the horizontal axis.}
\end{figure}

It is instructive to consider a simple two-body model with a core nucleus plus 
a valence neutron in order to understand the halo phenomenon. 
That is, we solve the Schr\"odinger equation,
\begin{equation}
\left[-\frac{\hbar^2}{2m}\,\frac{d^2}{dr^2} 
+\frac{l(l+1)\hbar^2}{2m r^2}+V(r)-\epsilon_l\right]
u_l(r)=0,
\end{equation}
where $m$ is the nucleon mass, $l$ is the single-particle angular momentum, 
and $V(r)$ is a spherical single-particle potential. 
Fig. 3 shows the radial wave function $u_l(r)$ on the linear 
(the upper panel) and the logarithmic (the lower panel) scales for 
the $^{11}$Be nucleus with the two-body $^{10}$Be+$n$ model. 
We use a Woods-Saxon potential for $V(r)$ with the radius and the 
diffuseness parameters of $R$=2.74 fm and $a$=0.75 fm, respectively, 
whereas the depth parameter is adjusted for each angular momentum $l$ 
so that the single-particle energy $\epsilon$ is $-$0.5 MeV. 
For simplicity, we do not consider a spin-orbit interaction. 
The solid, the dashed, and the dotted lines show the wave functions 
for the 2$s$, 1$p$, and 1$d$ states, respectively. 
One can clearly see that the wave function for the 2$s$ state is largely 
extended, while that for the 1$d$ state is spatially 
rather compact. The root-mean-square (rms) radii are 7.17, 5.17, and 4.15 fm 
for the 2$s$, 1$p$, and 1$d$ states, respectively. 
Fig. 4 shows the rms radii as a function of the single-particle energy. 
As one can show analytically\cite{RJM92}, 
the rms radius diverges for $l$ = 0 and 1 (it behaves as 
$|\epsilon|^{-1/2}$ for $l = 0$ and 
$|\epsilon|^{-1/4}$ for $l = 1$), while 
it converges to a constant value for higher values of $l$ in the limit of 
zero binding energy. 
The halo structure, therefore, has been ascribed to 
an occupation of a weakly-bound $l=0$ or $l=1$ orbit by a valence
nucleon near the threshold\cite{RJM92,S92}.

\subsection{Coulomb dissociation}

The halo structure significantly affects 
the dissociation process of a one-neutron halo 
nucleus in an external Coulomb field. 
The photoabsorption cross section for a dipole photon is given by 
\begin{equation}
\sigma_\gamma
=\frac{16\pi^3}{9\hbar c}
E_\gamma\cdot\frac{dB({\rm E1})}{dE_\gamma},
\label{sigma_gamma}
\end{equation}
where $E_\gamma$ is the photon energy and 
\begin{equation}
\frac{dB({\rm E1})}{dE_\gamma}
=
\frac{1}{2j_i+1}
\left|\langle \psi_f||e_{\rm E1}rY_{1}||\psi_i\rangle\right|^2
\delta(\epsilon_f-\epsilon_i- E_\gamma),
\label{BE1}
\end{equation}
is the reduced transition probability 
(see {\it e.g.,} Appendix B of Ref. \citen{RS80}). 
Here, $\psi_i$ and $\psi_f$ denote the wave functions for 
the initial and the final states, 
respectively, $j_i$ being the angular momentum for the initial state 
$\psi_i$. 
$e_{\rm E1}$ 
is the E1 effective charge, which is given by 
\begin{equation}
e_{\rm E1}=\frac{Z_1A_2-Z_2A_1}{A_1+A_2}\,e, 
\end{equation}
for a two-body system with a $(A_1,Z_1) + (A_2,Z_2)$ configuration.  

\begin{figure}[t]
\centerline{\psfig{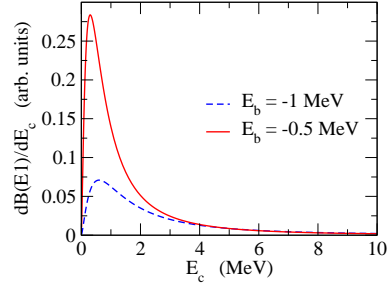}}
\caption{The $B$(E1) distribution from a bound $s$-wave state to continuum $p$-wave states 
given by Eq. 
(\ref{E1analytic}) for two different values of the binding energy, $E_b$. It is plotted 
as a function of the relative energy of the final continuum state, $E_c$. }
\end{figure}

A characteristic feature of the dipole excitation is that the $B$(E1) 
distribution, $dB({\rm E1})/dE$, has a strong peak in the low energy region 
when the binding energy is small (that is, the soft E1 mode\cite{HJ87,I87}). 
This large concentration of the E1 strength 
near the continuum threshold is caused by the optimal
matching of wave functions between a weakly bound and continuum
states.
For a transition from an $s$-wave state to a $p$-wave state, 
the $B$(E1) distribution can be evaluated analytically as
\begin{equation}
\frac{dB({\rm E1})}{dE}
=
\frac{3\hbar^2}{\pi^2\mu}\,e^2_{\rm E1}\, 
\frac{\sqrt{|E_b|}\,E_c^{3/2}}{(|E_b|+E_c)^4},
\label{E1analytic}
\end{equation}
if one employs 
a Yukawa function for $\psi_i$ and a spherical Bessel function for 
the radial part of $\psi_f$ \cite{BW52,BBH91,SGTIY95,OIF94} 
(see Refs. \citen{NLV05} and \citen{TB05} 
for a general expression with arbitrary initial and final angular momenta). 
Here, $\mu$ is the reduced mass between the two fragments, $E_b$ is the energy of the 
initial bound state, and $E_c$ is the energy of the final bound state, that is, 
the photon energy is $E_\gamma=E_c-E_b=E_c+|E_b|$. 
This equation indicates that the peak energy appears 
at $E_c=3|E_b|/5$ with the height of the peak being proportional to $1/|E_b|^2$. 
The total strength is given by 
\begin{equation}
B({\rm E1})=\int dE\,\frac{dB({\rm E1})}{dE}=\frac{3\hbar^2e^2_{\rm E1}}{16\pi^2\mu|E_b|}.
\end{equation}
Therefore, the peak position moves towards low energy as the binding energy, $|E_b|$, 
decreases, and at the same time the height of the peak increases, leading also 
to the increase of the total E1 strength. These features can be clearly seen in Fig. 5, 
which shows the $B$(E1) distribution 
given by Eq. (\ref{E1analytic}) for two different binding energies.  

Notice that using Eq. (\ref{BE1}) it is easy to derive that the total E1 strength 
(that is, the non-energy weighted sum rule) is proportional to the expectation value of 
$r^2$ with respect to the ground state, 
\begin{equation}
B({\rm E1})=\frac{3}{4\pi}\,e_{\rm E1}^2\langle r^2 \rangle_i.
\label{totE1}
\end{equation}
As we mention in the previous subsection, the rms radius diverges in the zero binding 
limit for $l$=0 and 1 states, leading therefore to a divergence of the total E1 strength. 
Thus, an observation of a large E1 strength makes a clear indication of a halo structure 
of the nucleus. 

\begin{figure}[t]
\centerline{\psfig{file=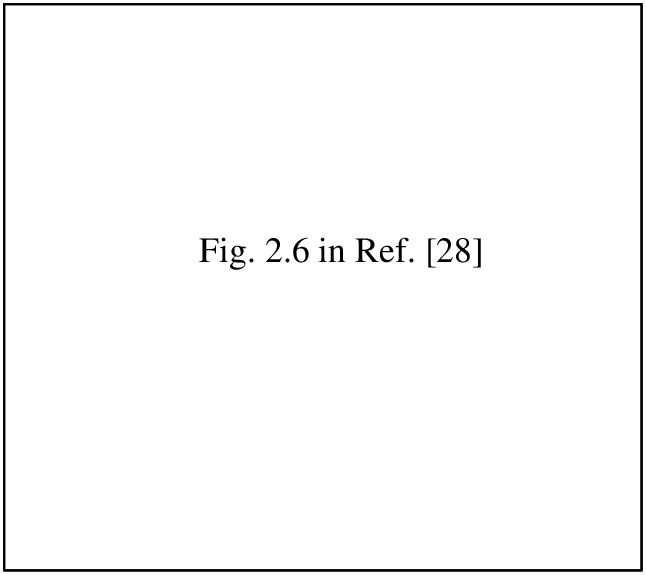,width=6.5cm}}
\caption{The experimental B(E1) distribution 
for the $^{11}$Be nucleus deduced from the Coulomb breakup with a $^{208}$Pb 
target at 70 MeV/nucleon\cite{N94}. 
Taken from Ref. \citen{NK12}.} 
\end{figure}

Experimentally, the Coulomb dissociation of halo nuclei has been studied by 
Coulomb excitation experiments with a heavy target nucleus, such as $^{208}$Pb
\cite{NK12}. 
The Coulomb breakup cross sections are often analyzed by the virtual photon 
theory, in which the cross sections are given as a product of the photo absorption 
cross sections, Eq. (\ref{sigma_gamma}), and the virtual photon 
flux, $N_{\rm E1}(E_\gamma)$\cite{BB88,WA79}. 
Figure 6 shows the experimental $B$(E1) distribution for the $^{11}$Be nucleus 
obtained by 
the Coulomb breakup reaction 
at 72 MeV/nucleon\cite{N94}. 
The observed dipole strength shows a strong peak at about 800 keV excitation energy, 
which is consistent with the binding energy of about 500 keV. The peak is 
large, and is again consistent with the halo structure of this nucleus. 

In addition to the large interaction cross sections\cite{Ozawa00,Takechi12}, 
large Coulomb breakup cross sections are experimentally observed also 
for $^{19}$C 
and $^{31}$Ne \cite{N99,N09}. These nuclei are thus considered to be good 
candidates for halo nuclei. 


\section{Two-neutron halo nuclei}

\subsection{Two-nucleon correlation}

Let us now discuss properties of two-neutron halo nuclei, in which 
two valence neutrons are weakly bound to a core nucleus. 
For these nuclei, one must consider a (pairing) interaction between 
the valence neutrons. 
It has been well recognized that 
the pairing correlation plays an important role in nuclear 
physics\cite{RS80,DNW96,BHR03,BB05}. 
It leads to an extra binding for even-mass nuclei, and at the same time 
reduces the level 
density in the low energy region. Also, a pairing interaction scatters 
nucleon pairs from a single-particle level below the Fermi surface to 
those above, and consequently each single-particle level
is occupied only partially. 
For weakly bound nuclei, the pairing interaction 
works 
by scattering nucleon pairs inevitably to unbound states. 

If one adopts a three-body model for a two-neutron halo nucleus, 
one could view it as a system of two interacting Fermions inside a 
confining potential. This is in a sense similar to a 
problem of interacting Fermion gas in a harmonic trap in atomic physics 
(see {\it e.g.,} Refs. \citen{CB07} and \citen{EKLN11}, and references 
therein). 
But a problem is much more challenging in weakly bound nuclei, because 
a trapping potential is not an infinite well (in contrast to a harmonic trap) 
so that the couplings to 
continuum are important, and also because a trapping potential itself 
is constructed self-consistently from the interaction among nucleons. 

Among two-neutron halo nuclei, the so called Borromean nuclei have 
attracted lots of attention. 
These are unique three-body bound systems, in which
any two-body subsystem is not bound \cite{BE91,Zhukov93}.
Typical examples include $^{11}$Li and $^6$He,
which can be viewed as three-body systems consisting
of a core nucleus and two valence neutrons.
Since both the $n$-$n$ and $n$-core two-body subsystems are not bound,
these nuclei are bound only as three-body systems.

One of the most important current open questions concerning the Borromean
nuclei is to clarify the characteristic nature of correlations between
the two valence neutrons, which do not form a bound state in the vacuum.
For instance, a spatial structure of two valence neutrons 
in the Borromean nuclei has attracted much attention. As a matter of fact, 
this has a long history of research 
as a general problem in nuclear 
physics. One of the oldest publications on this problem is by 
Bertsch, Broglia, and Riedel, who solved a shell model for $^{210}$Pb 
and showed that the two valence neutrons 
are strongly clusterized \cite{BBR67}. 
Subsequently, Migdal argued that two neutrons may 
be bound in a nucleus even though 
they are not bound in the vacuum \cite{M73}. 
The strong localization of two neutrons inside a nucleus has been 
referred to as {\it dineutron correlation}. 
It has been shown in Ref. \citen{CIMV84} that the dineutron 
correlation is caused by admixtures of a few single-particle orbits with 
opposite parity. 

\begin{figure}[t]
\psfig{file=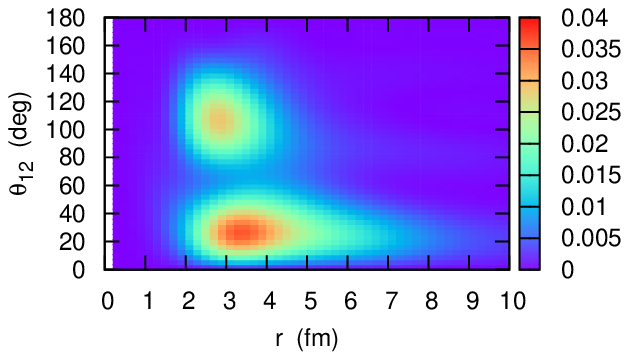,width=5.5cm}
\psfig{file=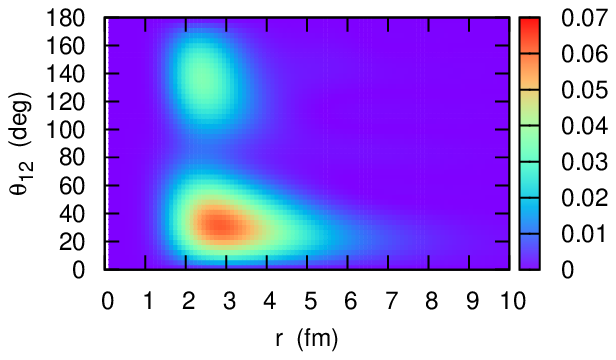,width=5.5cm}
\caption{
The two-particle densities for $^{11}$Li (the left panel) and for $^6$He (the 
right panel) obtained with a three-body model calculation \cite{HS05}. 
These are plotted as a function of neutron-core distance, $r_1=r_2\equiv r$, 
and the opening angle between the valence neutrons, $\theta_{12}$. 
The densities are weighted with a factor $8\pi^2r^4\sin\theta_{12}$. } 
\end{figure}

Although the dineutron correlation exists even in stable nuclei
\cite{IAVF77,PSS07,HSS10}, it is enhanced in weakly bound nuclei 
because the admixtures of single-particle orbits with different 
parities are easier due to the couplings to the continuum spectra. 
Probably it was Hansen and Jonson who
exploited the idea of dineutron correlation explicitly for
exotic nuclei for the first time.
They proposed the
dineutron cluster model and successfully analyzed the matter radius
of $^{11}$Li \cite{HJ87}. They also predicted a large Coulomb dissociation
cross section of the $^{11}$Li nucleus.
In the 1990's, more microscopic three-body model calculations 
for neutron-rich nuclei started\cite{BE91,Zhukov93}. 
These three-body model calculations have revealed that 
a strong dineutron correlation, where the two valence neutrons take
a spatially compact configuration, indeed exists in weakly-bound 
Borromean nuclei \cite{IMKT10,BE91,Zhukov93,BBBCV01,HS05,HSCS07,KKM10}. 
It has been shown that the dineutron correlation exists also in heavier 
neutron-rich nuclei \cite{HTS08,IIAAK08,E07,MMS05,PSS07,AI09}
as well as in infinite neutron matter \cite{M06,MSH07,EHSS09}.
The diproton correlation, which is a counter part of the dineutron 
correlation, has also been shown to 
exist  in the proton-rich 
Borromean nucleus, $^{17}$Ne\cite{OHS10}. 

Figs. 7(a) and 7(b) show the two-particle density 
obtained with three-body model calculations 
for $^{11}$Li and $^6$He, respectively. 
These are plotted as a function of the neutron-core distance,
$r_1=r_2\equiv r$ and the opening angle between the valence neutrons,
$\theta_{12}$. A weight of
$4\pi r^2\cdot 2\pi r^2 \sin \theta_{12}$ has been multiplied.
See Ref. \citen{HS05} for the details
of the calculations.
One can see that a large fraction of two-particle density is
concentrated in the region with small opening angle $\theta_{12}$ for 
both the nuclei. 
This is a clear manifestation of the 
strong dineutron correlation discussed in this subsection. 

\subsection{Coulomb breakup}

\begin{figure}[t]
\psfig{file=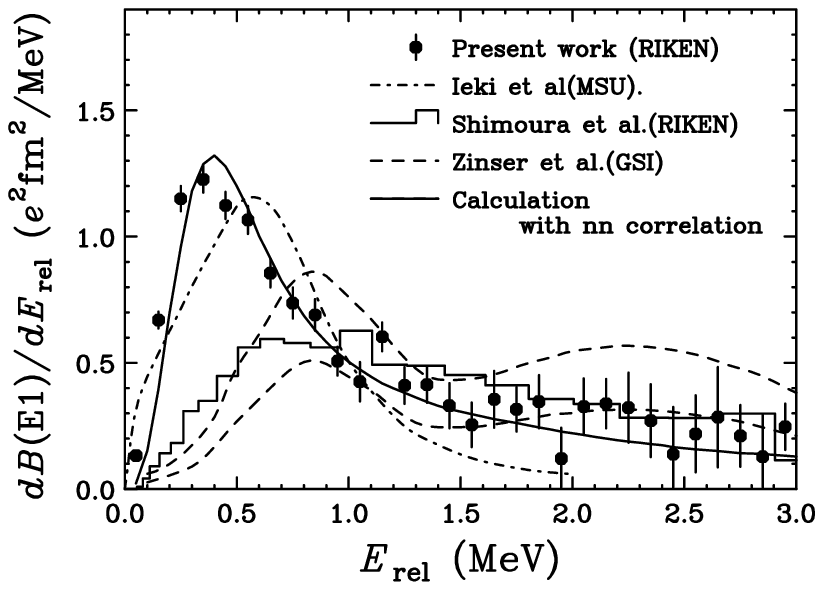,width=5.5cm}
\psfig{file=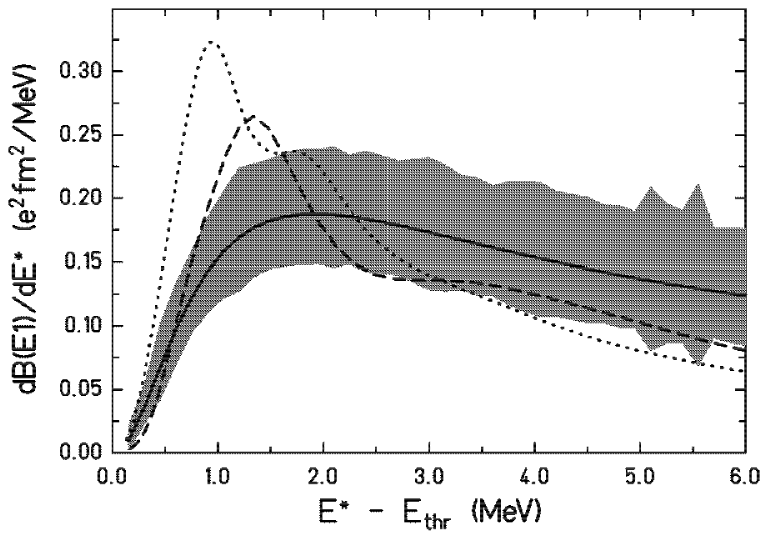,width=5.5cm}
\caption{
The experimental B(E1) distributions for the $^{11}$Li (the left panel) 
and for the $^6$He (the right panel) nuclei deduced from the Coulomb 
breakup measurements. Taken from Refs. \citen{N06} and \citen{A99}. }
\end{figure}

The Coulomb breakup of two-neutron halo nuclei can be discussed 
in a similar manner as that of one-neutron halo nuclei discussed in 
Sec. 2.2. 
The only difference is that the E1 operator is now given by 
\begin{equation}
\hat{D}_\mu = e_{\rm E1}\,RY_{1\mu}(\hat{\vec{R}}),
\label{E1}
\end{equation}
where 
\begin{equation}
\vec{R}=\frac{\vec{r}_1+\vec{r}_2}{2},
\end{equation}
is the center of mass coordinate for the two valence neutrons, and the 
E1 effective charge is given by
\begin{equation}
e_{\rm E1}=\frac{2Z_c}{A_c+2}\,e,
\end{equation}
with $A_c$ and $Z_c$ being the mass and charge numbers for the core 
nucleus. 

The left and the right panels of Fig. 8 
show the measured $B$(E1) distribution for the $^{11}$Li 
and $^6$He nuclei\cite{N06,A99}, respectively. 
Those $B$(E1) distributions, especially that for the $^{11}$Li nucleus, show 
a strong concentration in the low excitation region, 
similar to the $B$(E1) distribution for the $^{11}$Be nucleus shown in Fig. 6, 
reflecting the halo structure of these nuclei. 
Moreover, the experimental data for $^{11}$Li are consistent with the 
theoretical calculation only when the interaction between the valence 
neutrons is taken into account, strongly 
suggesting the existence of the dineutron correlation in 
this nucleus (see also Ref. \citen{EHMS07}). 

\begin{figure}[t]
\psfig{file=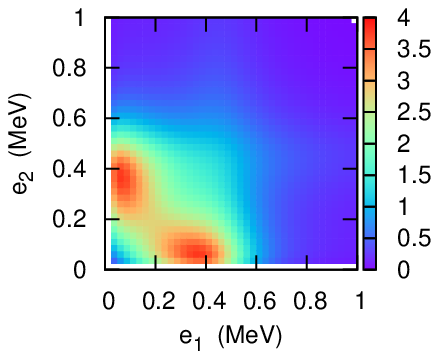,width=5.5cm}
\psfig{file=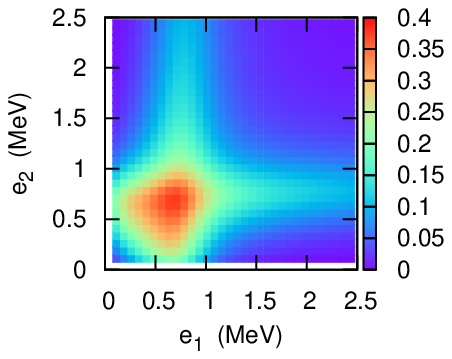,width=5.5cm}
\caption{
The dipole strength distribution, $d^2B({\rm E1})/de_1de_2$, of
$^{11}$Li (the left panel) and $^6$He (the right panel) 
as a function of
the energies of the two emitted neutrons relative to the core nucleus.
It is plotted in units of $e^2$fm$^2$/MeV$^2$. 
See Ref. \citen{HSNS09} for the details of the calculations.}
\end{figure}

The left and the right panels of Fig. 9 show calculated 
dipole strength distributions, $d^2B({\rm E1})/de_1de_2$, of $^{11}$Li and $^6$He, 
respectively, obtained with the three-body model\cite{HSNS09} 
together with the Green's function method for the continuum dipole 
response\cite{EB92}. 
Here, $e_1$ ($e_2$) is the relative energy between
the first (second) neutron and the core nucleus.
One immediately notices that the strength distribution is considerably
different between $^{11}$Li and $^6$He, despite similar ground state 
density to each other (see Fig. 7). 
This difference has been shown to be due to the 
different resonance properties of the neutron-core interaction between 
the two nuclei\cite{HSNS09} (see also Ref. \citen{KKM10}).

\subsection{Charge radii of halo nuclei} 

Until recently Na was the lightest element of which 
charge radii (or proton density distribution radii) of neutron-rich 
short lived isotopes have been measured.  
Combined with 
the measurements of matter 
radii, the development of neutron skins 
thus has been presented for Na isotopes\cite{OST01}. 
For neutron halo nuclei, interaction cross section and fragmentation 
measurements have provided a  mean to determine the matter density distribution.   
However the proton density distribution could not be 
determined directly from such measurements.   Although 
the proton distribution 
has been considered no to extend out even when neutron halos are formed, 
based on 
the fact that narrow momentum distributions, 
that indicate long tails of distribution, 
are observed only for a neutron or two-neutron removal channels 
of fragmentation, no direct determination of proton 
distribution radii (or charge radii) was possible until recently.

Recent developments of ion traps now provide a means to determine 
charge radii of very light neutron rich nuclei including neutron halo 
nuclei, $^6$He, $^{11}$Li, and $^{11}$Be.  
The charge radii are determined by the isotope shift measurements 
of atomic transitions.  In a very light atom, 
the isotope shift $\delta \nu$ includes two terms, 
\begin{equation}
\delta\nu=\delta\nu_{\rm MS}+\delta\nu_{\rm FS}.
\end{equation}
The first term $\delta\nu_{\rm MS}$ is the mass shift that is proportional 
to the difference of the masses ($A$ and $A'$) of two isotopes, 
\begin{equation}
\delta\nu_{\rm MS}\propto \frac{A-A'}{AA'},
\end{equation}
and $\delta\nu_{\rm FS}$  is the field shift which is proportional 
to the difference of the rms radii of nuclei,
\begin{equation}
\delta\nu_{\rm FS}\propto Z\times\Delta[\Psi(0)]^2\times\delta\langle 
r^2\rangle,
\end{equation}
where $Z$ is the atomic number of the isotopes and 
$\Delta[\Psi(0)]^2$ 
is the difference of the electron wave function at the nuclei.  
To obtain the field shift, the mass shift has to be calculated 
theoretically and then subtracted from the total isotope shift 
determined from the measurement.

For a light nucleus as an example, the mass shift term for $^6$He and $^4$He 
is more than 10$^5$ times larger than the shift expected from a change of 
radius. The mass shift term cannot be separately measured experimentally 
so that an accurate theoretical estimation of this term is necessary 
in order to determine the charge radii. The recent development of atomic theory 
of a few electron system has enabled to do such a calculations 
successfully up to three electron systems\cite{YD03}. 

The charge radii of $^6$He and $^8$He have been determined by the 
ANL group using ANL/ATLAS and GANIL RIB facilities to be 
2.054$\pm$0.014 fm and 1.93$\pm$0.03 fm, respectively \cite{Wan04,Mue07}.  
Those experiments use a magneto-optical trap of atoms for precision 
laser spectroscopy. 
The charge radii of Li and Be isotopes are  determined by a GSI group 
using TRIUMF/ISAC facility and ISOLDE RIB facilities \cite{San06,Nor09}.  
In these experiments, a Doppler-free two-photon transition was used 
for the lithium measurements \cite{San06,Ewa04,Nor11}  
and collinear laser spectroscopy with a frequency 
comb for the beryllium isotopes \cite{Nor09}. 


\begin{figure}[t]
\psfig{file=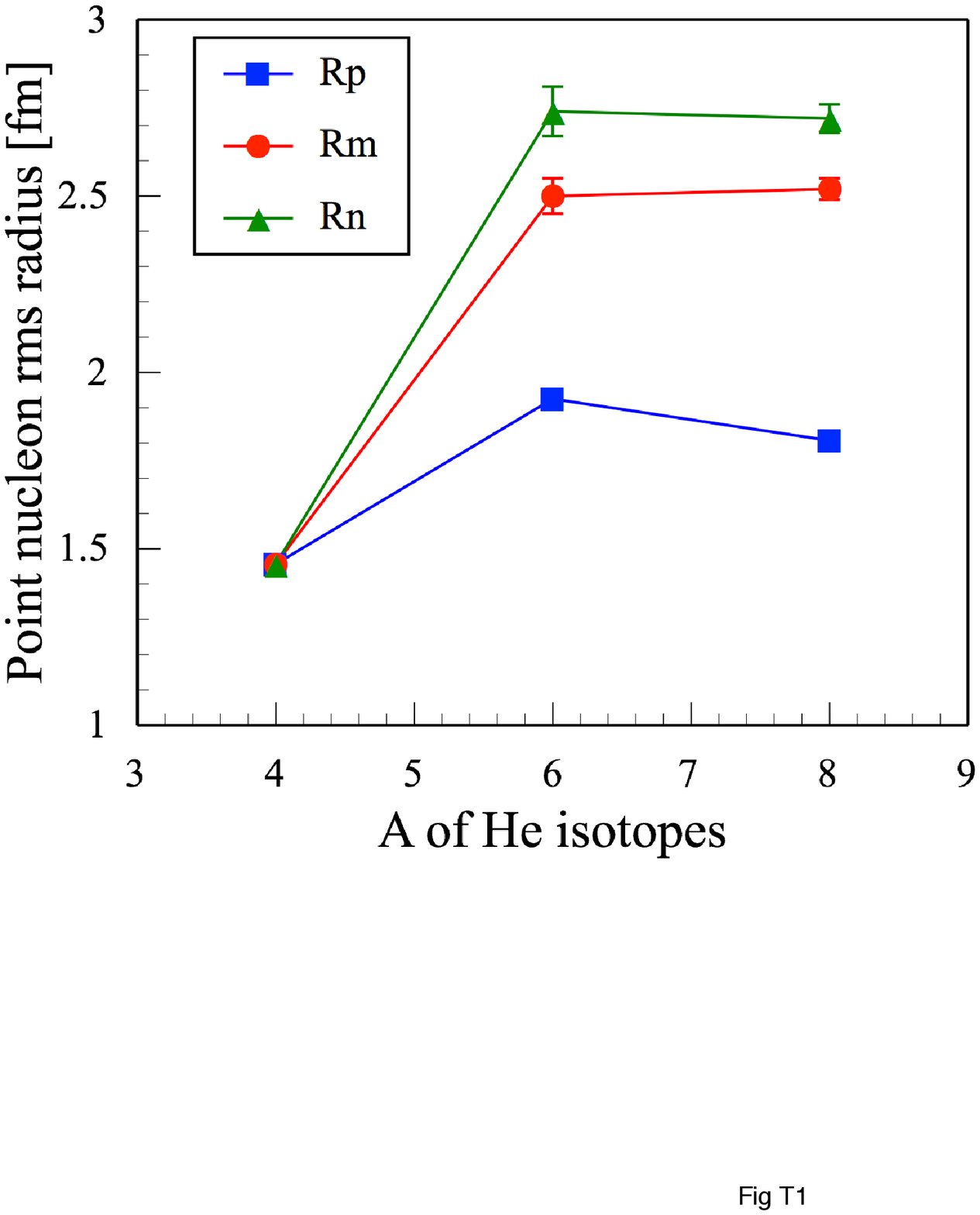,clip,width=5cm}
\psfig{file=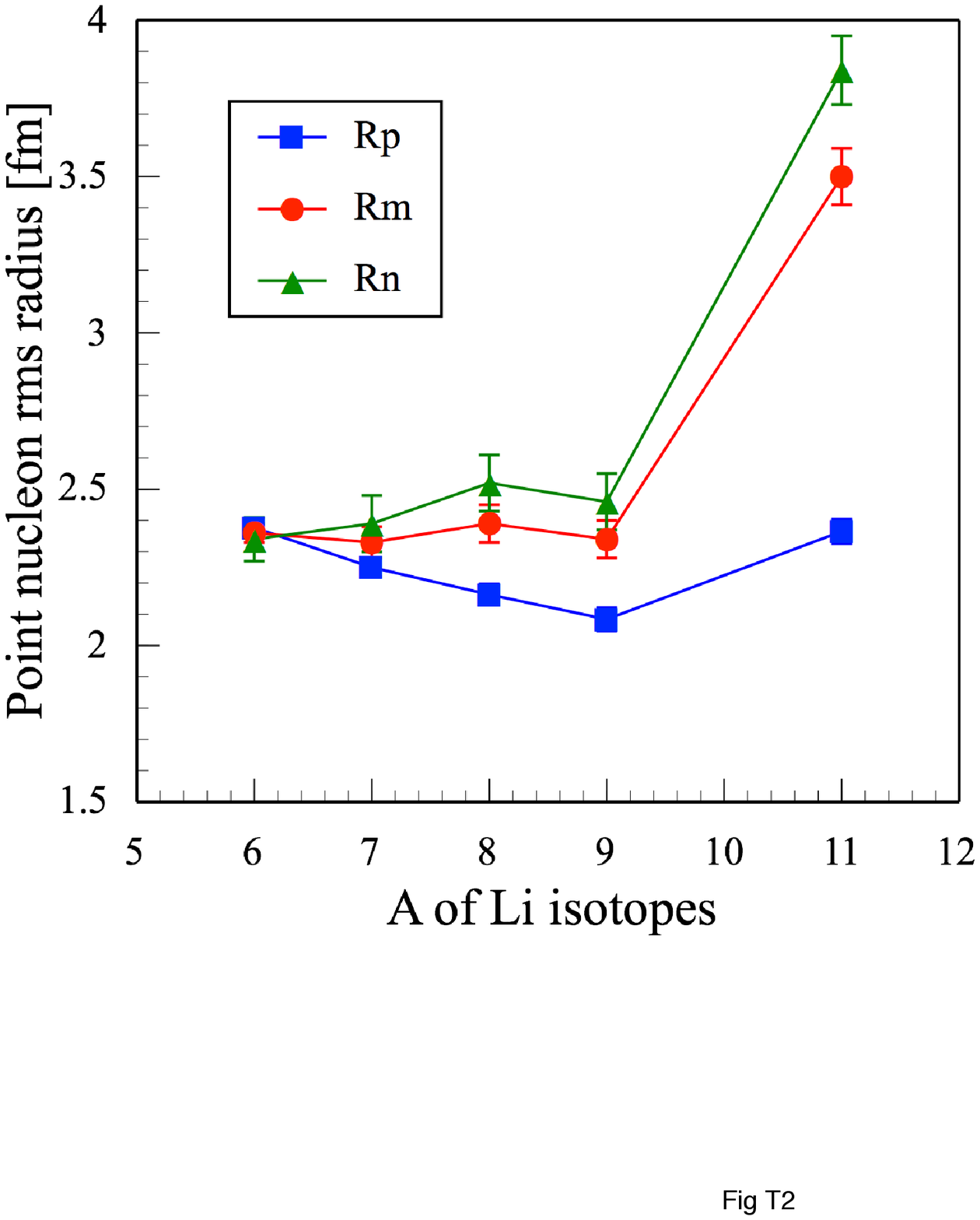,clip,width=5cm}
\psfig{file=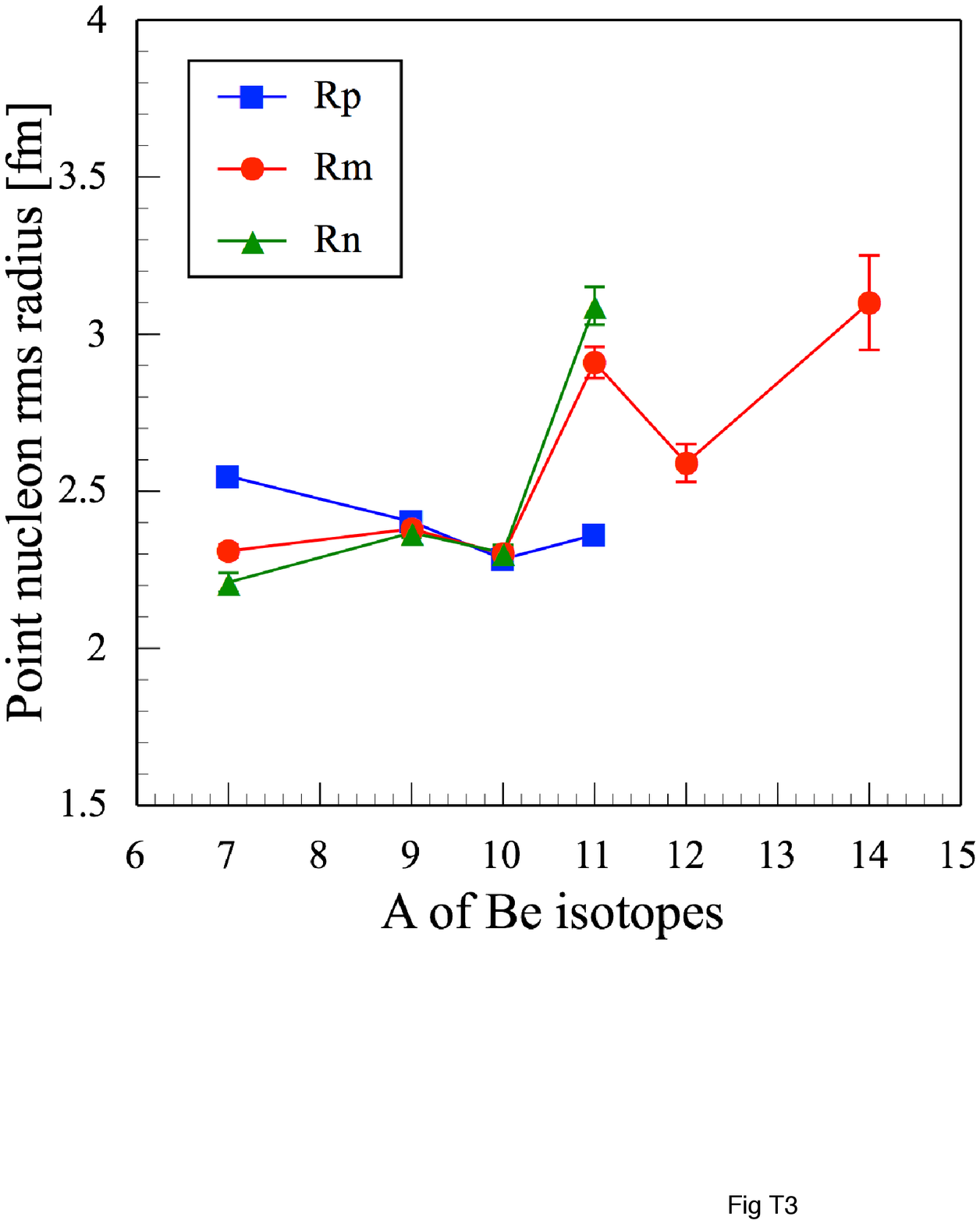,clip,width=5cm}
\caption{The matter, proton, and neutron rms radii ($R_m$, $R_p$, and $R_n$) 
for He, Li, and Be isotopes determined from the interaction/reaction 
cross section 
and the charge radius measurements. 
}
\end{figure}

The proton distribution rms radii, $R_p\equiv \sqrt{\langle r_p^2\rangle}$, 
has been calculated from the charge radii, $\sqrt{\langle r_{ch}^2\rangle}$, 
by,
\begin{equation}
\langle r_{ch}^2\rangle =  \langle r_p^2\rangle +
\langle \rho_p^2\rangle +
\frac{N}{Z} \langle \rho_n^2\rangle +\frac{3\hbar^2}{4m_p^2c^2},
\end{equation}
where $r_p$ is the radius of point proton distribution of a 
nucleus, $\rho_p$ and $\rho_n$ are the charge radii of free proton 
and free neutron,  $\langle \rho_p^2\rangle =0.769\pm 0.012$ 
and $\langle \rho_n^2\rangle =-0.1161\pm 0.0022$ fm$^2$ \cite{Yao06}, 
and the last term is the so called Darwin-Foldy term (0.033 fm$^2$) 
\cite{Fri97}.
The relation between matter, proton, and neutron rms radii is written as,
\begin{equation}
AR_m^2=ZR_p^2+NR_n^2.
\label{rmsradius}
\end{equation} 
The rms radii of matter (or point nucleon distribution), $R_m$, have been 
determined by 
the interaction cross section and reaction cross section measurements 
\cite{OST01}.  The matter ($R_m$), proton ($R_p$), 
and neutron ($R_n$) radii so determined are presented in Fig. 10 for 
He, Li, and Be isotopes. 
One can clearly see that $R_p$ is much smaller that that of neutron $R_n$ 
in neutron rich isotopes and in particular in halo nuclei. 
However the proton radius increases slightly when neutron halo is formed. 
This observation is consistent with the view of a core+decoupled halo 
neutron(s) structure of halo nuclei that has been widely used for modeling 
halo nuclei. Under this model, the core of a halo nucleus has the same 
proton distribution as the isolated core nucleus. Because of the large 
distance between the core and the halo neutron(s), 
the core moves around the center of mass of the halo nucleus and 
therefore the proton radius in a halo nucleus is larger than that 
of isolated core.

\subsection{Geometry of two-neutron halo nuclei} 

Under the assumption of core+ two-neutron for a two-neutron 
halo nucleus, the spatial correlation of halo neutrons can be studied. 
The geometry of the model for such a nucleus is shown in Fig. 11. 

\begin{figure}[t]
\centerline{\psfig{file=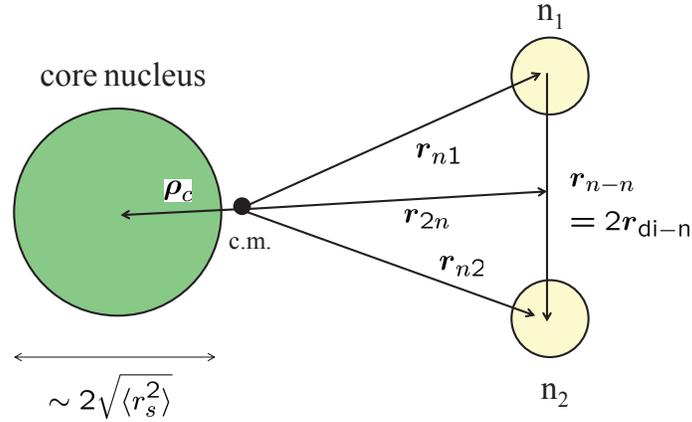,clip,width=10cm}}
\caption{The geometry of the three-body model for two-neutron 
halo nuclei. 
}
\end{figure}

The relation between nucleon, proton, and neutron mean-square (ms) radii 
for a nucleus $i$ 
is given by a similar equation as Eq. (\ref{rmsradius}).  
Therefore from the measurements of matter and charge radii of a halo 
nucleus and its core nucleus, all radii of the halo nucleus, 
$\langle r_{m}^2\rangle$, $\langle r_{p}^2\rangle$, 
and $\langle r_{n}^2\rangle$, 
and the core, 
$\langle r_{sm}^2\rangle$, $\langle r_{sp}^2\rangle$, 
and $\langle r_{sn}^2\rangle$, 
are determined.
One can also define 
the ms matter radii of the halo neutrons, $\langle r_h^2\rangle$, 
and that of the core, $\langle r_{cm}^2\rangle$, which are related as,
\begin{equation}
A\langle r_{m}^2\rangle = A_c\langle r_{cm}^2\rangle
+A_h\langle r_{h}^2\rangle,
\label{geo}
\end{equation}
where $A$, $A_c$, and $A_h$ are the mass numbers of the halo nucleus, 
the core nucleus, and the number of halo neutrons, respectively.  

Using the coordinate of the core nucleus, $\vec{\rho}_c$, relative to the 
center of mass of the halo 
nucleus,  that is, 
the movement of the center of the core in 
the halo nucleus,  the relation between 
the nucleon, proton, and neutron radii between the core and the halo 
nucleus are represented as,
\begin{eqnarray}
\langle r_{cp}^2\rangle &=& 
\langle r_{p}^2\rangle =
\langle r_{sp}^2\rangle +\langle \rho_{c}^2\rangle, \\
\langle r_{cm}^2\rangle &=& \langle r_{sm}^2\rangle +\langle \rho_{c}^2\rangle, 
\label{rcm}
\\
\langle r_{cn}^2\rangle &=& \langle r_{sn}^2\rangle +\langle \rho_{c}^2\rangle, 
\end{eqnarray}
From those equations, one can obtain the movement of the core, 
$\langle \rho_c^2\rangle$.  
Then the ms radius of the halo distribution, $\langle r_h^2\rangle$, 
is also determined from Eq. (\ref{geo}). 

The distance between the center of mass of the halo nucleus 
and the center of mass of the two halo neutrons ($r_{2n}$) 
can be calculated from $\rho_c$,
\begin{equation}
A_h^2\langle r_{2n}^2\rangle = A_c^2\langle \rho_c^2\rangle.
\label{r2n}
\end{equation}
The distance between the core and the two-neutron center of mass 
is $\sqrt{\langle r^2_{c-2n}\rangle}=\sqrt{\langle r^2_{2n}\rangle}
+\sqrt{\langle \rho^2_{c}\rangle}$. 
Let us define the size of di-neutron as the distribution 
radius of the neutrons around the center of mass of halo neutrons, 
$\vec{r}_{n-n}=\vec{r}_{n1}-\vec{r}_{n2}=2\vec{r}_{di-n}$, and thus 
$r^2_{n-n}=4r^2_{di-n}$, that is, 
$r_{di-n}$ is the radius of the di-neutron forming the halo.
The radius of the halo distribution in the halo nucleus 
and the dineutron radius is related as,
\begin{equation}
\langle r_h^2\rangle = \langle r^2_{2n}\rangle +
\langle r^2_{di-n}\rangle.
\label{rh}
\end{equation}
Using these $\langle r_h^2\rangle$ and 
$\langle r_{2n}^2\rangle$, 
one can determine the di-neutron ms radius and thus the ms 
separation distance $\langle r^2_{n-n}\rangle$ of the two halo neutrons.
Notice that combining Eqs. (\ref{geo}), (\ref{rcm}), (\ref{r2n}), and 
(\ref{rh}) yields
\begin{equation}
\langle r_m^2\rangle = \frac{A_c}{A_c+2}\langle r_{sm}^2\rangle 
+\frac{1}{A_c+2}\left(\frac{2A_c}{A_c+2}\langle r_{c-2n}^2\rangle +
\frac{1}{2}\langle r_{n-n}^2\rangle \right),
\end{equation}
for $A_h=2$. 

One can also obtain the two-neutron cross term 
$\langle \vec{r}_{n1}\cdot \vec{r}_{n2}\rangle$ as 
\begin{equation}
\langle \vec{r}_{n1}\cdot \vec{r}_{n2} \rangle= \frac{1}{4}
\left(A_c^2\langle \rho_c^2\rangle-\langle r_{n-n}^2\rangle\right), 
\end{equation}
from
\begin{equation}
A_c^2\langle \rho_c^2\rangle=
\langle (\vec{r}_{n1}+\vec{r}_{n2})^2\rangle,~~~~{\rm and}~~~
\langle r_{n-n}^2\rangle = 
\langle (\vec{r}_{n1}-\vec{r}_{n2})^2\rangle,
\end{equation}
where $A_h=2$ has been used. 

\begin{table}[hbt]
\tbl{The root-mean-square (rms) radii of $^6$He. 
$\sqrt{\langle r_m^2\rangle}$, $\sqrt{\langle r_p^2\rangle}$, 
and $\sqrt{\langle r_n^2\rangle}$ are the point nucleon, 
proton, and neutron rms radii of the nucleus, respectively. 
$\sqrt{\langle r_h^2\rangle}$ is the rms radius of the halo-neutron 
distribution. 
$\sqrt{\langle r_{2n}^2\rangle}$ is the rms distance from the center 
of mass of the nucleus to the center of mass of the two valence 
neutrons, while 
$\sqrt{\langle r_{c-2n}^2\rangle}$ is the rms distance from the $^4$He core 
to the center of mass of the two valence neutrons. The first 
estimation for the latter quantity by Wang {\it et al.} 
was 3.71$\pm$0.07 fm \cite{Wan04}. 
$\sqrt{\langle r_{n-n}^2\rangle}$ is the rms distance between the 
valence neutrons, while 
$\langle \vec{r}_{n1}\cdot\vec{r}_{n2}\rangle$ is the correlation of 
the two valence neutrons. All these radii are given in the 
unit of fm, except for 
$\langle \vec{r}_{n1}\cdot\vec{r}_{n2}\rangle$, which is given in the 
unit of fm$^2$. 
See text for explanation. 
GFMC, NCSM, and AMD denote Greens Function Monte Carlo, No-Core Shell Model, 
and Anti-symmetrized Molecular Dynamics, respectively. 
}
{
\begin{tabular}{c|c|c|c|c|c|c|c|c} 
\hline
& Experiment & GFMC\cite{EHMS07} 
& Varga\cite{VSO94} & Esbensen\cite{EBH97}
& Funada\cite{FKS94} & Zhukov \cite{Zhukov93}& NCSM \cite{CN06}& AMD \cite{K07}\\
\hline
$\sqrt{\langle r_m^2\rangle}$ 
& 2.43$\pm$ 0.03 & & 2.46 & & & 2.45 & & 2.23 \\
$\sqrt{\langle r_p^2\rangle}$ 
& 1.912$\pm$ 0.018 & & 1.80 & & & & 1.89$\pm$0.04 & 1.83 \\
$\sqrt{\langle r_n^2\rangle}$ 
& 2.65$\pm$ 0.04 & & 2.67 & & & & 2.67$\pm$0.05 & 2.40 \\
$\sqrt{\langle r_n^2\rangle}
-\sqrt{\langle r_p^2\rangle}$ 
& 0.808$\pm$ 0.047 & & 0.87 & & & & &  \\
$\sqrt{\langle r_h^2\rangle}$
& 3.37$\pm$ 0.11 & & & & & & &  \\
$\sqrt{\langle r_{2n}^2\rangle}$
& 2.52$\pm$ 0.05 & & 3.42& & & & &  \\
$\sqrt{\langle r_{c-2n}^2\rangle}$
& 3.84$\pm$ 0.06 & 3.81$\pm$0.20& & 3.63 & 3.51& 3.54& &  \\
$\sqrt{\langle r_{n-n}^2\rangle}$
& 3.93$\pm$ 0.25 & & & 4.62 & 4.55& 4.58& &  \\
$\langle \vec{r}_{n1}\cdot\vec{r}_{n2}\rangle$ (fm$^2$)
& 2.70$\pm$ 0.97 & & & 0.54 & 0.292& 0.325& &  \\
\hline
\end{tabular}}
\end{table}

\begin{table}[hbt]
\tbl{Same as Table I, but for $^{11}$Li. 
The first estimation of 
$\sqrt{\langle r_{c-2n}^2\rangle}$ by Sanchez {\it et al.} 
was 6.2$\pm$0.3 fm\cite{San06}. SHF and TOSM denote Skyrme Hartree-Fock 
and Tensor Optimized Shell Model, respectively. }
{
\begin{tabular}{c|c|c|c|c|c} 
\hline
& Experiment & Esbensen\cite{EBH97} &  SHF \cite{SR96}
& Zhukov \cite{Zhukov93}& TOSM \cite{MKKTI08}\\
\hline
$\sqrt{\langle r_m^2\rangle}$ 
& 3.50$\pm$ 0.09 & & 2.87 & 3.39 & 3.41 \\
$\sqrt{\langle r_p^2\rangle}$ 
& 2.37$\pm$ 0.04 & & 2.28 & & 2.34 \\ 
$\sqrt{\langle r_n^2\rangle}$ 
& 3.84 $\pm$ 0.11 & & 3.06 & & 3.73 \\ 
$\sqrt{\langle r_n^2\rangle}
-\sqrt{\langle r_p^2\rangle}$ 
& 1.48$\pm$ 0.12 & & & & \\
$\sqrt{\langle r_h^2\rangle}$
& 6.1$\pm$ 0.3 & & & &  \\
$\sqrt{\langle r_{2n}^2\rangle}$
& 5.0$\pm$ 0.5 & & & &  \\
$\sqrt{\langle r_{c-2n}^2\rangle}$
& 6.2$\pm$ 0.5 & 5.12 & 6.26& 5.69  \\
$\sqrt{\langle r_{n-n}^2\rangle}$
& 7.0$\pm$ 1.7 & 6.77& & & 7.33  \\
$\langle \vec{r}_{n1}\cdot\vec{r}_{n2}\rangle$ (fm$^2$)
& 2.70$\pm$ 0.97 & & & &  \\
\hline
\end{tabular}}
\end{table}

The empirical values for those variables for $^6$He and $^{11}$Li, 
extracted from the experimental interaction cross sections and charge 
radii,  
are 
shown in Table 1 and 2.  Corresponding theoretical values are 
also shown in the tables for a few model calculations.
The data for $^6$He and $^{11}$Li show that the distance between 
the core and the halo neutrons, $\sqrt{\langle r_{c-2n}^2\rangle}$, 
is almost equal to the distance between two neutrons, 
$\sqrt{\langle r_{n-n}^2\rangle}$, indicating that two-neutrons 
are sitting close together and have strong {\it di-neutron} correlations.
The opening angle 
between the valence neutrons 
with respect to the core nucleus 
can be calculated from the empirical values for 
$\sqrt{\langle r_{c-2n}^2\rangle}$ and 
$\sqrt{\langle r_{n-n}^2\rangle}$ to be 
$\langle \theta_{12}\rangle = 58.9\pm 12.6$ 
and 
$\langle \theta_{12}\rangle = 54.2\pm 3.04$ 
degrees for $^{11}$Li and $^6$He, respectively. 
(We should remark
here that it is misleading to say that two neutrons are mostly
sitting with opening angles obtained in this way. Instead, the
mean opening angle is an average of a smaller
and a larger correlation angles in the density distribution shown in Fig. 7.) 

In principle, 
a similar analysis can be made for $^8$He.  However it is known that 
the possible core $^6$He is known not to be a good inert core and thus 
$^6$He + 2$n$ model is very poor.  Instead, it is known from the 
fragmentation experiment \cite{Tan92a} that $^4$He + 4$n$ is a good 
model of $^8$He ground state.  Following the similar procedure 
as above, one can determine the movement of the 4-valence 
neutron center-of-mass in $^8$He, $\sqrt{\langle r^2_{4n}\rangle}$ to be 
1.07$\pm$0.05 fm. This value 
is much smaller than that in $^6$He, 
$\sqrt{\langle r^2_{2n}\rangle}/2 =1.62 \pm 0.03$ fm.  
It is an indication that the four neutrons in $^8$He are 
distributed more uniformly than in $^6$He.

Notice that the core-2n rms distance, $\sqrt{\langle R^2\rangle}=
\sqrt{\langle r_{c-2n}^2\rangle}$, for two-neutron halo nuclei 
can be estimated also 
from the Coulomb dissociation cross sections. 
For the transition operator given by Eq. (\ref{E1}), 
the expectation value of $R^2$ in the ground state 
can be estimated from the total $B$(E1) value as 
\begin{equation}
B_{\rm tot}({\rm E1})
\sim
\frac{3}{\pi}\left(\frac{Z_ce}{A_c+2}\right)^2\,\langle R^2\rangle,
\end{equation}
(see also Eq. (\ref{totE1})). 
Using the experimental 
matter radii to estimate the rms distance $\sqrt{\langle r_{n-n}^2\rangle}$, 
the opening angle 
between the valence neutrons 
with respect to the core nucleus 
has been extracted as \cite{HS07} 
$\langle \theta_{12}\rangle = 65.2\pm 12.2$ degrees for 
$^{11}$Li and $74.5\pm 12.1$ degrees for $^6$He 
(see also Ref. \citen{BH07}), which agree well with the results of the 
three-body model calculation of Ref. \citen{HS05}. 
The value of 
$\langle \theta_{12}\rangle$  for $^{11}$Li
is consistent also with 
the one obtained in Table 1 and 2, although  
the value for $^6$He is somewhat larger. 

\subsection{Two-nucleon radioactivity}

Although
the experimental observation of
the strong low-lying dipole strength distribution 
in the $^{11}$Li nucleus (see Fig. 8) 
has provided an
experimental signature of
the existence of dineutron correlation in this nucleus, 
it is still an open question how to probe it directly. 
That is, in the Coulomb breakup process, 
the ground state wave function of a two-neutron halo nucleus is 
perturbed by the external electromagnetic field of the 
target nucleus, and it may not be easy to disentangle the 
dineutron correlation 
in the ground state from that in the excited states. 
It would be desirable if one could find an observable which 
reflects only the properties of the ground state. 

\begin{figure}[t]
\centerline{\psfig{file=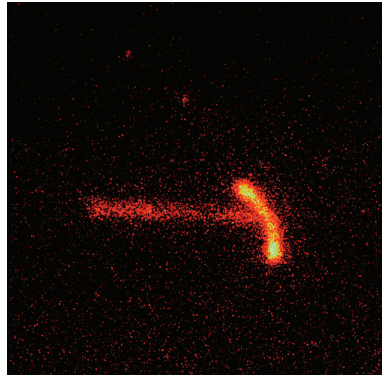,width=5cm}}
\caption{
An example of trajectories of the two emitted protons from the 
two-proton radioactivity of $^{45}$Fe recorded by a CCD camera. 
Taken from Ref. \citen{MDJ07}.}
\end{figure}

The two-proton radioactivity, that is, a spontaneous emission of 
two valence protons, of proton-rich nuclei 
\cite{PKGR12,WD97,BP08,G09,GWM09} 
is expected to provide a good tool to probe the di-proton 
correlation in the initial wave function. 
Nuclei beyond the proton drip line are unstable against proton emission,
but, since a proton has to penetrate the Coulomb barrier, their
lifetime is sufficiently long to study their spectroscopic
properties. A single-proton radioactivity has been found in many odd-Z 
proton-rich nuclei, and has provided a powerful tool to study 
the spectroscopy of proton-rich nuclei beyond the proton-drip line 
\cite{PKGR12,WD97,SDW99,KGB03,TBR06}. 
When the single-proton emission is energetically forbidden, 
proton-rich nuclei beyond the proton drip line decay via emission of 
two protons. Even though this process had been predicted 
theoretically in 1960\cite{G60}, its first experimental discovery was 
much later, only in 2002 \cite{PBB02,GBC02}, for the $^{45}$Fe nucleus. 
Subsequently, the energy and the angular distributions of the two emitted 
protons were also measured \cite{MDJ07}. 
An impressive development in Ref. \citen{MDJ07} 
was the use of a new type of a gaseous detector, 
in which images of ionizing particle trajectories 
can be optically recorded with a CCD camera. 
An example of recorded trajectories of the two emitted protons 
from the two-proton radioactivity of $^{54}$Fe is shown in Fig. 12. 
This technique was used also for the two-proton radioactivity of 
$^{48}$Ni\cite{PPD11}. 

\begin{figure}[t]
\centerline{\psfig{file=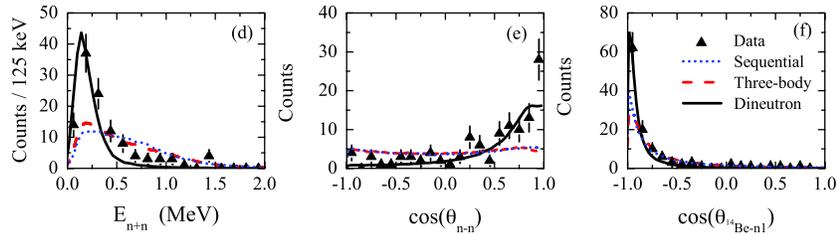,width=11cm}}
\caption{
The energy and the angular distributions of the two emitted neutrons 
from the two-neutron radioactivity of $^{16}$Be. 
Taken from Ref. \citen{SKB12}.}
\end{figure}

Very recently, a ground state {\it two-neutron} emission 
was discovered for the first time for $^{16}$Be\cite{SKB12}. 
This is an analogous process of the two-proton radioactivity, 
corresponding to a penetration of two neutrons over a centrifugal 
barrier. 
For the $^{16}$Be nucleus, the one-neutron emission process is 
energetically forbidden, which makes $^{16}$Be an ideal two-neutron 
emitter. 
It is remarkable that 
the observed energy and the angular distributions of the 
two emitted neutrons show a strong indication of the dineutron 
correlation in the ground state of $^{16}$Be (see Fig. 13)\cite{SKB12}. 

\subsection{Two-neutron transfer reactions}

It has been recognized for a long time 
that two-neutron transfer reactions 
are sensitive to the pairing correlation\cite{Y62,OV01,PGB11,GLV12}. 
The probability for the two-neutron transfer process is enhanced 
as compared to a naive expectation of sequential transfer process, 
that is, the square of one-neutron transfer probability\cite{OBG87,CSP11}. 
The enhancement of pair transfer probability has been attributed to 
the pairing effect, such as 
the enhancement of pair strength function \cite{KSGG04,SM11}
and the surface localization of a Cooper 
pair\cite{CIMV84,ILM89}. 
The pair transfer reaction is thus considered to provide 
a promising way to probe
the dineutron correlation discussed in Sec. 3.1. 
However, the reaction dynamics is rather complicated 
and has not even been well established. 
For instance, it is only in a recent calculation 
that a theoretical calculation achieves a satisfactory agreement 
with the experimental data\cite{PBM11}. 
It would therefore be not surprising that 
the role of dineutron in the pair transfer reaction
has not yet been fully clarified. 

One example is a relative importance of the one-step (the simultaneous 
pair transfer) process and the two-step (the sequential pair transfer) 
process. In heavy-ion pair transfer reactions of stable nuclei, both  
processes are known to play a role \cite{BC82,FBLP78,EJR98}. 
For weakly-bound nuclei, most of the intermediate states for the two-step 
process are likely in the continuum spectra. 
It is still an open question how 
this fact, together with the $Q$-value matching condition \cite{B72}, 
alters the dynamics of the pair transfer reaction of neutron-rich 
nuclei\cite{VS12}. 

\begin{figure}[t]
\centerline{\psfig{file=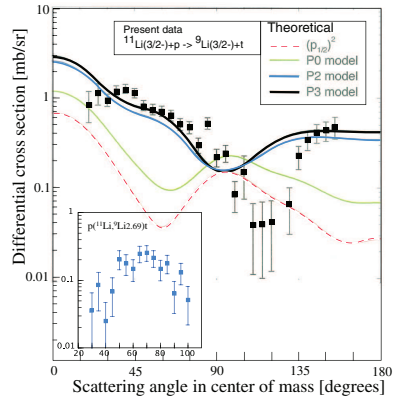,clip,width=5.5cm}}
\caption{
The experimental angular distributions for 
the $^1$H($^{11}$Li,$^9$Li)$^3$H reaction at 3 MeV/nucleon.
Taken from Ref. \citen{TAB08}. }
\end{figure}

On the other hand, 
the cross sections for the pair transfer reaction of the Borromean 
nuclei, $^{11}$Li and $^6$He, have been measured recently
\cite{TAB08,Oga99,Oga99a,Raa99,Gio05,CNS08}. 
The data for the $^1$H($^{11}$Li,$^9$Li)$^3$H reaction at 3 MeV/nucleon 
indicate that the cross sections are indeed sensitive to the pair 
correlation in the ground state of $^{11}$Li (see Fig. 14) \cite{TAB08}. 
That is, the experimental cross sections can be accounted for 
only when the $s$-wave component is mixed in the ground state 
of $^{11}$Li by 30-50\%. 
Another important finding in this measurement is that significant 
cross sections were observed for the pair transfer process to 
the first excited state of $^9$Li\cite{TAB08}, 
which has made a good support for 
the idea of phonon mediated pairing mechanism\cite{PBVB10}.

\begin{figure}[t]
\centerline{\psfig{file=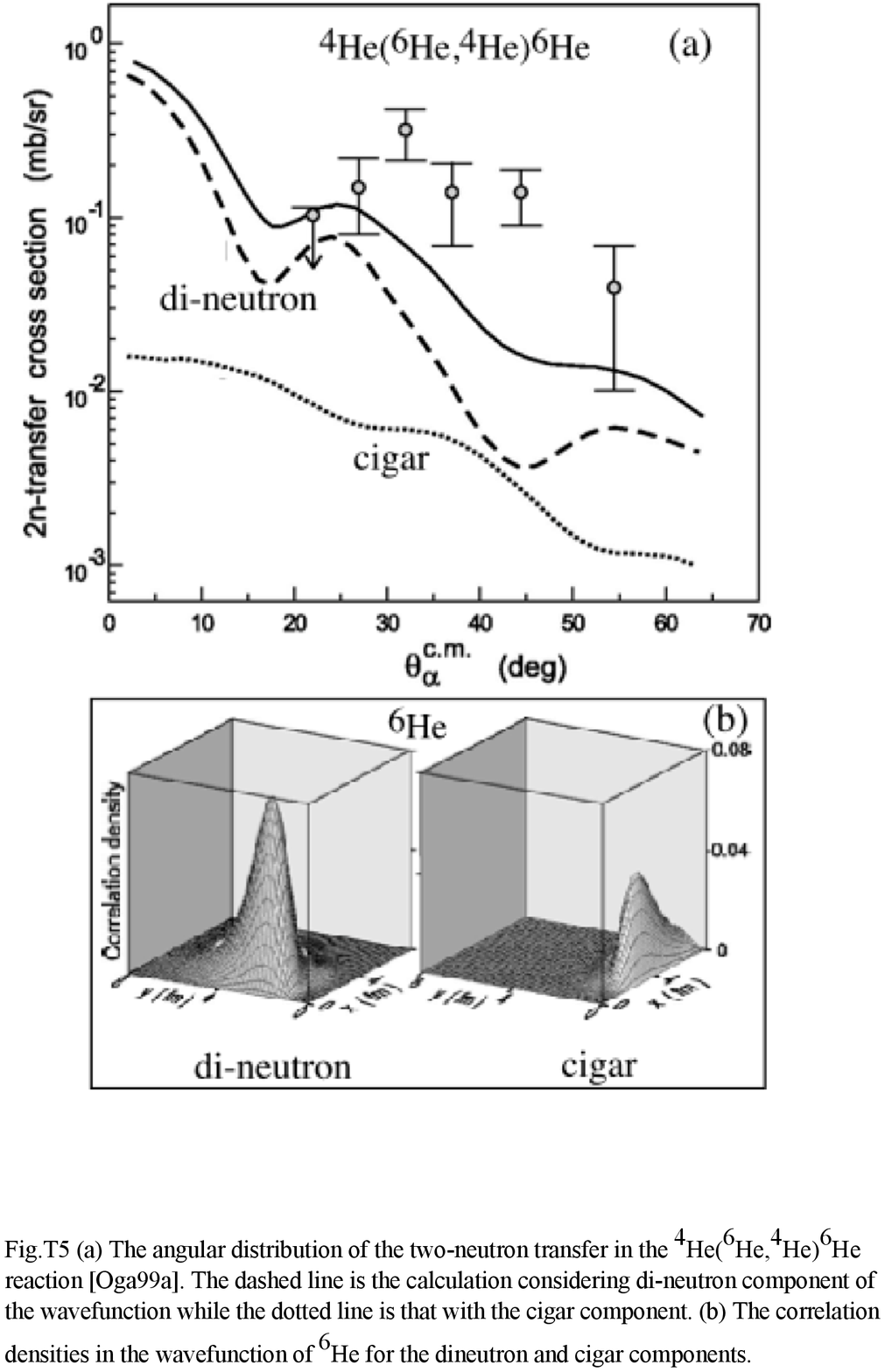,clip,width=7cm}}
\caption{(a) 
The angular distribution of the two-neutron transfer 
in the $^4$He($^6$He,$^4$He)$^6$He reaction \cite{Oga99a}. 
The dashed line shows the calculation considering the dineutron 
component of the wave function while the dotted line is that with the 
cigar component. (b) The correlation densities in the wave function of $^6$He 
for the dineutron and the cigar components. 
}
\end{figure}

The two-neutron transfer from $^6$He was investigated using $^4$He and $p$ 
targets at FLNR(Dubna) at 151 MeV \cite{Oga99, Oga99a}.  
The angular distribution from the $^4$He($^6$He,$^4$He)$^6$He reaction 
shows dominant contribution from di-neutron configuration (see the dashed 
line in Fig. 15 (a)).  While the wavefunction contains both the cigar 
and di-neutron correlation the amplitude of the latter is larger 
(Fig.15 (b), see also Fig. 7).  
The elastic scattering $^4$He($^6$He,$^6$He)$^4$He was also studied 
at center of mass energies of 11.6 MeV and 15.9 MeV at the ARENAS facility 
at Louvain-La-Neuve \cite{Raa99}.  An interpretation of the data 
at both energies \cite{Oga99, Raa99} showed that the rise of cross section 
at large angle were due to two-neutron transfer.  The calculation 
was carried out based on the coupled reaction channels approach \cite{Kho04} 
including both one-step and sequential two-step transfer with realistic 
form factors.   It was further found \cite{Kho04} that the direct 
two-neutron transfer strongly dominates over the sequential transfer 
at the low energies where the minimum in the angular distribution 
\cite{Raa99, Kho04} is due to direct 2n transfer.  The sequential 
transfer becomes more sizable for the higher-energy data, 
though direct 2n transfer still dominates.  In the analysis in 
Ref. \citen{Kho04}, $^6$He was modeled as core +2$n$ bound state wavefunction 
where two $p_{3/2}$ neutrons coupled to $J$=0 (S-wave) and 1 (P-wave). 
The direct two-neutron transfer was found to be mainly due 
to the contribution from the S-wave.  The P-wave part of the 
wave function belongs to the cigar type configuration, which is 
found to be a smaller contribution.  The main conclusion is thus 
the same as Ref. \citen{Oga99}, and the dominance 
of S-wave two-neutron cluster transfer shows strongly correlated 
two neutrons in $^6$He.

\begin{figure}[t]
\centerline{\psfig{file=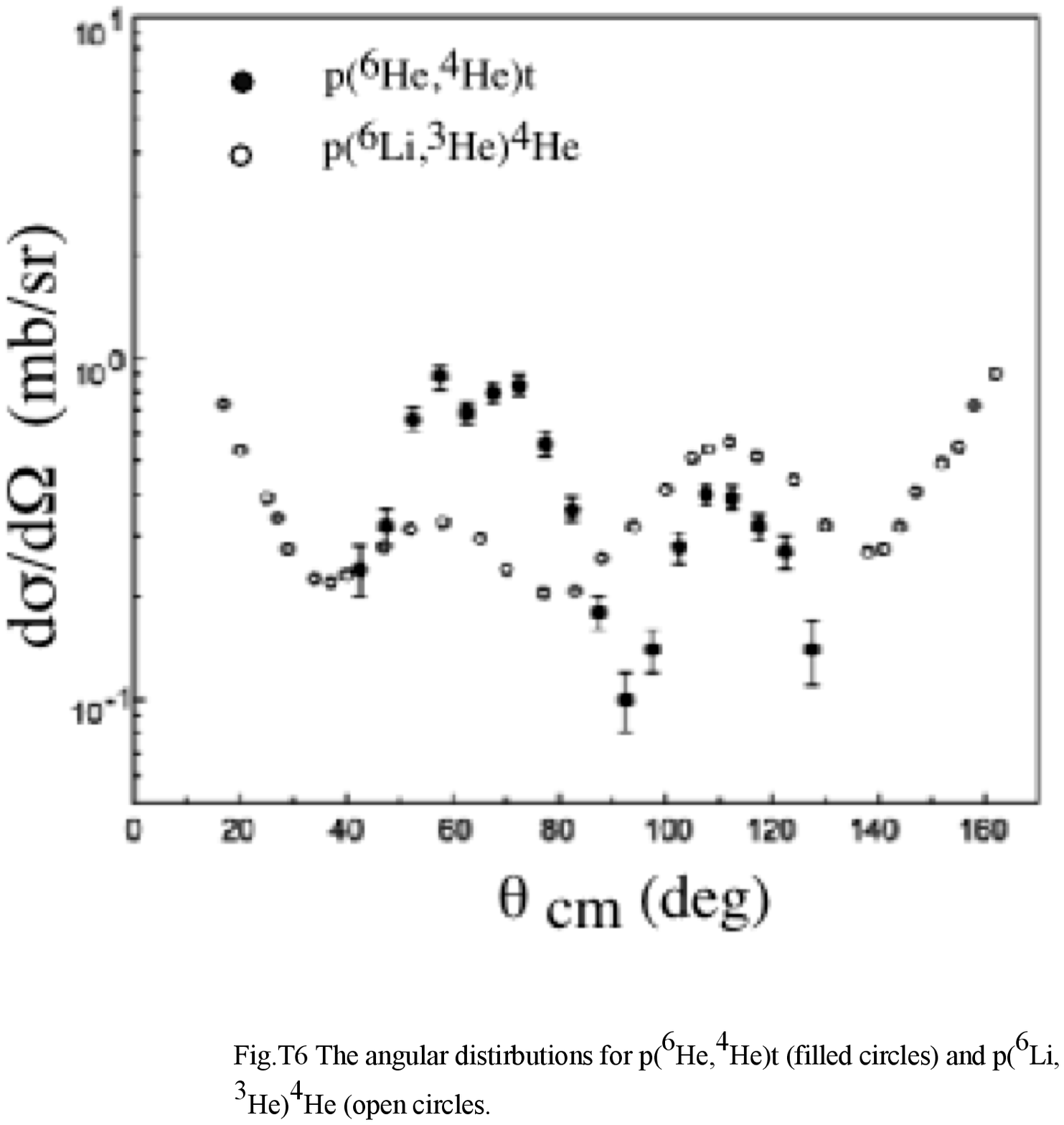,clip,width=7cm}}
\caption{
The angular distributions for $p(^6$He,$^4$He)$t$ (the filled circles) and 
$p(^6$Li,$^3$He)$^4$He (the open circles)\cite{Gio05}. 
}
\end{figure}

     The $p$($^6$He,$^4$He)$t$ angular distribution \cite{Oga99} (Fig. 16) 
when compared with the $^6$Li($p$,$^3$He)$^4$He, shows a slightly 
larger cross section for 2$n$ transfer at the center-of-mass (cm) 
scattering angles 
near 60 degrees. This was interpreted to be a signature of a more 
disperse 2$n$ wavefunction in $^6$He compared to the compact $d$ in $^6$Li.  
In both reactions a direct 2$n$ transfer was assumed to be the dominant 
process.  The $p(^6$He,$t$) reaction was also studied at 25 MeV/nucleon  
at GANIL \cite{Gio05}, with an aim at looking into possible $t$+$t$ 
cluster existing in $^6$He beside the interest in exploring the $^4$He+2$n$ 
configuration. The data are closer in magnitude and shape to the 2$n$ 
transfer than the transfer of tritons.  Therefore, the $^4$He+2$n$ 
configuration of $^6$He is again confirmed. 

In the experiment with a heavier target, that is, 
the $^6$He+$^{65}$Cu reaction at $E_{\rm lab}$=22.6 MeV, 
both the 1-neutron (1$n$) and the 2-neutron (2$n$) transfer cross sections were 
measured \cite{CNS08}. An interesting observation for this system 
is that the cross sections 
for the 2$n$ transfer are much larger than those for 
the 1$n$ transfer\cite{CNS08}. For stable nuclei, usually the 1$n$ transfer 
cross sections are much larger than the 2$n$ transfer cross 
sections\cite{OV01}, and the opposite observation for the $^6$He+$^{65}$Cu 
system can be regarded as 
a characteristic feature of a Borromean nucleus. 
A similar tendency, although less clearly, 
has been observed also for the transfer reactions for the 
$^8$He+$^{197}$Au system at a similar energy\cite{LNR11}. This measurement 
also shows that the transfer cross sections for the $^8$He projectile 
are considerably larger than those for the $^6$He projectile 
at energies around the Coulomb barrier\cite{LNR11}, while these nuclei 
behave similar to each other in the subbarrier fusion reactions\cite{LSN09}. 
Further theoretical studies are apparently necessary in order to understand 
the differences and the similarities of these 
Borromean nuclei, $^6$He and $^8$He, in several 
reaction processes at energies around the Coulomb barrier.

\section{Heavier neutron-rich nuclei}

\subsection{Matter radii and neutron skin thickness}

As we mentioned in Sec. 1, interaction cross sections $\sigma_I$ 
are intimately 
related to the size of colliding nuclei\cite{OST01,Ozawa00}. 
The interaction cross section is defined as a cross section for the  change 
of the proton number $Z$ and/or the neutron number $N$ 
of a projectile nucleus after the interaction with a target nucleus. 
Another important quantity is a reaction cross section $\sigma_R$, which is 
defined as 
the total cross section minus elastic cross section, 
that is, the sum of the interaction cross section and 
the inelastic scattering cross section. 
Interaction cross sections are easier to measure than reaction 
cross sections, while the opposite is the case for a theoretical 
evaluation. 
For neutron-rich nuclei, cross sections for inelastic
scattering are expected to be negligibly small at high incident energies 
\cite{Ozawa02,OYS92,KIO08},
and the interaction cross sections are almost the same
as the reaction cross sections.
Because of this, measured interaction cross sections have often 
been compared to calculated reaction cross sections. 

Reaction cross sections have often been analyzed by the Glauber 
theory\cite{AN03,OYS92,G59,OKYS01}. 
In the optical limit approximation to the Glauber theory, 
together with the zero range approximation to the nucleon-nucleon 
interaction, the reaction cross section is given by\cite{K75,BBS89}
\begin{equation}
\sigma_R=2\pi\int^\infty_0bdb\,\left[1-\exp\left(
-\sigma_{NN}\int d^2s\rho_P^{(z)}(\vec{s})\rho_T^{(z)}(\vec{s}-\vec{b})\right)
\right],
\end{equation}
where $\vec{b}$ is the impact parameter 
and $\vec{s}=(x,y)$ is the plane perpendicular to $z$. 
$\sigma_{NN}$ is the total $NN$ cross section, and $\rho^{(z)}(\vec{s})$  
is defined by $\rho^{(z)}(\vec{s})=\int dz \rho(\vec{r})$, $\rho_P$ and 
$\rho_T$ being the projectile and the target densities, respectively. 
It has been known that the optical limit approximation
overestimates reaction cross sections for weakly-bound nuclei
\cite{BES90,TUKS92,AKT96,AKTT96,AIS00}.
In order to cure this problem, 
Al-Khalili and Tostevin developed a few-body treatment for the 
Glauber theory\cite{AKTT96}. 
Abu-Ibrahim and Suzuki have also proposed another 
simple method which effectively takes into account the higher order 
corrections\cite{AIS00}. 

\begin{figure}[t]
\psfig{file=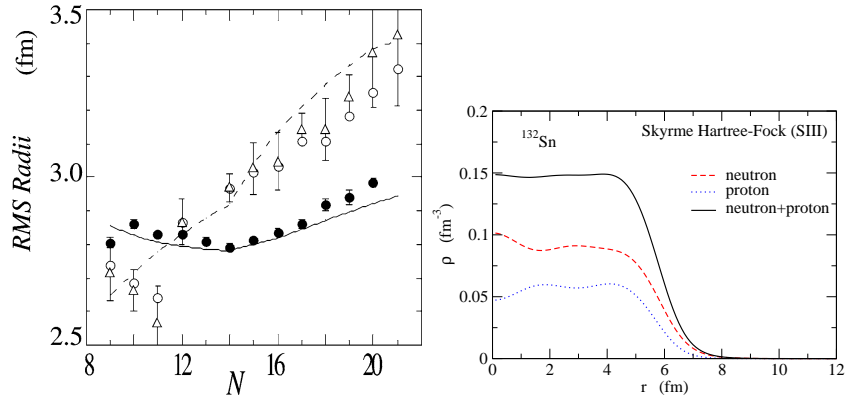,clip,width=5.5cm}
\psfig{file=fig17b.eps,clip,width=5.5cm}
\caption{
The left panel: Neutron 
rms radii for Na isotopes (the open symbols). These are deduced from 
the measured interaction cross sections using the proton rms radii 
(the filled circles) obtained with the isotope shift measurements. 
Taken from Ref. \citen{OST01}. 
The right panel: the density distribution for $^{132}$Sn obtained with 
the Skyrme-Hartree-Fock method with the SIII parameter set. 
}
\end{figure}

The left panel of Fig. 17 shows the neutron rms radii for Na isotopes (the open 
symbols)\cite{OST01} deduced from the measured interaction 
cross sections together 
with the proton rms radii (the filled circles) 
obtained from the isotope shift measurements. 
The figure indicates that the neutron rms radii are significantly larger 
than the proton rms radii for neutron-rich Na isotopes. 
This suggests that the neutron density distributions are largely extended 
over the proton density distributions, despite that there is no clear 
separation between a core nucleus and valence neutrons as in halo nuclei 
discussed in Sec. 2 and Sec. 3
(see Refs.\citen{RD09,RBD09,MR98,MTZ02,TZZM06,GYSG06} for the halo structure 
in heavy nuclei). This structure is referred to as 
skin structure\cite{MS80,FOT93,MDL00}. 
It should be mentioned that the skin structure is not necessarily 
related to $s$- and $p$- wave single-particle orbits, in contrast to 
the halo structure\cite{FOT93}. That is, the skin structure can be 
realized even with higher angular momentum states. 
As an example of skin nucleus, the right panel of Fig. 17
shows the density 
distribution for $^{132}$Sn obtained with the Skyrme-Hartree-Fock 
method\cite{VB72} with the SIII parameter set\cite{SIII}. 

As we have discussed in Sec. 3.3, 
in order to discuss the skin thickness of neutron-rich nuclei, 
one needs both the matter and charge radii. 
For Na isotopes shown in Fig. 17, the proton radii have been 
obtained with the isotope shift measurements. 
The isotope shift measurement is not always 
applicable, however. 
In that case, one may use the charge changing cross section 
to estimate the rms radii for the proton 
distribution\cite{BGG92,CBC00,BCE98,YFF10,YHK11}. 
This is the cross section for a change in the charge number $Z$, and 
is considered to be sensitive to the proton distribution\cite{MZT02,BG04}. 
Alternatively, one may use the proton elastic scattering 
measurement\cite{TST08,ZSM10}. 
A yet novel method to extract the information on the proton distribution 
of neutron-rich nuclei 
is to use the electron scattering. 
Using a self-containing RI target (SCRIT), an attempt has 
already been successfully 
commenced at RIBF at RIKEN, Japan\cite{WEF08,SWE09}. 

\subsection{Odd-even staggering of interaction cross sections}

\begin{figure}[t]
\psfig{file=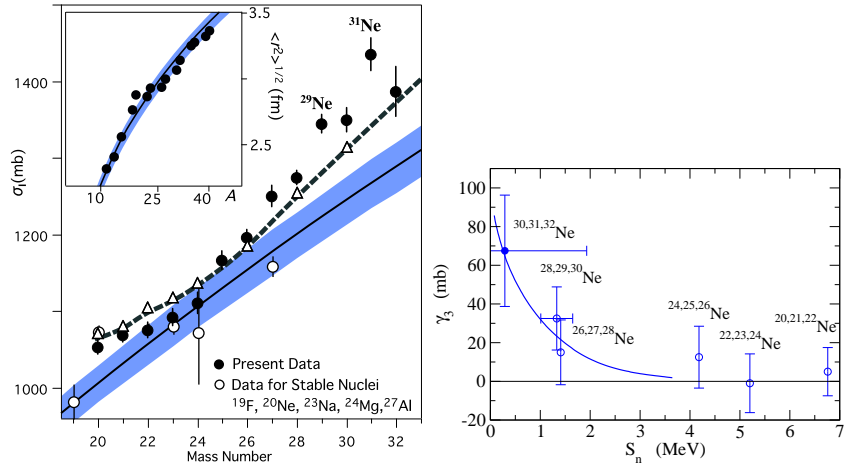,clip,width=5.5cm}
\psfig{file=fig18b.eps,clip,width=5.5cm}
\caption{
The left panel: 
The experimental interaction cross sections for Ne isotopes with a carbon 
target at around 
240 MeV/nucleon. 
The inset shows rms radii of 
stable nuclei. 
Taken from Ref. \citen{Takechi12}. 
The right panel: the odd-even staggering parameter defined by Eq. (\ref{oes}) 
for Ne isotopes as a function of the one-neutron separation energy for 
the odd-mass nuclei\cite{HS12}. 
The solid line shows the result of the Hartree-Fock-Bogoliubov 
calculations. 
}
\end{figure}

The experimental interaction cross sections for neutron-rich nuclei
often show a large odd-even staggering (OES). That is,
the cross section for an odd-mass nucleus is significantly larger than
the cross sections for the neighboring even-mass nuclei.
A typical example is the interaction cross sections for $^{30,31,32}$Ne,
measured recently by Takechi {\it et al.} (see the left panel of 
Fig. 18) \cite{Takechi12}.
In Ref. \citen{HS12}, the odd-even staggering was analyzed 
by introducing the staggering parameter defined as 
\begin{equation}
\gamma_3=(-)^{A}\frac{\sigma_I(A+1)-2\sigma_I(A)+\sigma_I(A-1)}{2},
\label{oes}
\end{equation}
where $\sigma_I(A)$ is the interaction cross section of a nucleus
with mass number $A$. The right panel of Fig. 18 shows 
the experimental staggering parameter $\gamma_3$ for Ne isotopes 
as a function of 
the neutron separation energy for the
odd-mass nuclei. 
It is compared with the results of the Hartree-Fock-Bogoliubov (HFB) 
calculations that takes into account the pairing correlations 
in the mean-field approximation. 
One can clearly see that the staggering parameter $\gamma_3$
increases rapidly for small separation energies, and goes up to
a large value reaching $\gamma_3\sim$ 80 mb.
Also, the experimental staggering parameters agree well with the HFB 
calculations, suggesting that the pairing correlation plays an important 
role in the odd-even staggering of the interaction cross sections. 

\subsection{Alpha cluster in neutron-rich nuclei}

The $\alpha$ cluster model has been successful in describing 
the structure of light $N=Z$ nuclei\cite{PTPSuppl52,PTPSuppl68}. 
In neutron-rich nuclei, extra neutrons are surrounding 
tightly-bound alpha particles. 
A new theoretical framework, the antisymmetrized molecular dynamics (AMD), 
has been developed by Horiuchi {\it et al.}\cite{OHMO92}, which has been 
successfully applied to neutron-rich nuclei\cite{KEH01,KEK10,KEKO12}. 
In this method, a many-body wave function is assumed to be a parity- and 
angular momentum projected Slater 
determinant with multi-centered Gaussian single-particle wave functions.  
Generator coordinate method (GCM) calculations with such Slater determinants 
have also been considered\cite{KH04}. 
In this method, the alpha cluster 
is not assumed a priori, in contrast to the conventional cluster model, 
but it can emerge as a result of the energy minimization. 
The AMD calculations show that the cluster structure is developed 
and stabilized in some neutron-rich nuclei\cite{KEH01}, 
which can be well interpreted in terms 
of the molecular-orbital picture\cite{IOI01,IOIO04}.  

\subsection{Shell evolution: 
change of spherical magic numbers in neutron-rich nuclei}

\begin{figure}[t]
\centerline{\psfig{file=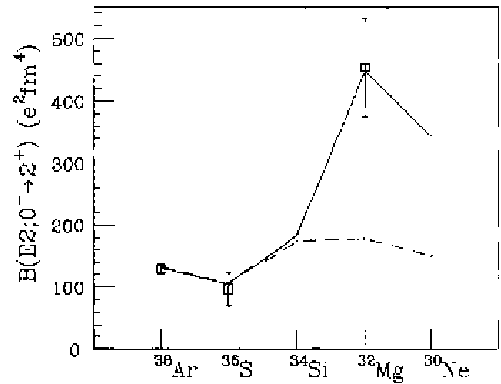,width=7cm}}
\caption{
The measured $B({\rm E2})$ values for $N=20$ even-even nuclei. 
The solid and the dashed lines are the results of the shell model 
calculation with the model space of sd+pf shells and sd shells, respectively. 
Taken from Ref. \citen{OF96}. }
\end{figure}

The shell closures and the associated magic numbers are one of the 
most important concepts in nuclear physics\cite{MJ49}. 
For stable nuclei, these magic numbers correspond to 2, 8, 20, 28, 50, 82, 
and 126. That is, if either the neutron number or the proton number 
(or both) coincides with one of these numbers, that 
nucleus is particularly rigid and takes a spherical shape. In general, the 
first 2$^+$ state has a relatively large excitation energy in nuclei at the 
shell closures. However, a $\beta$-decay spectroscopy studies had 
revealed by the middle of 1980's 
that the first 2$^+$ state of the neutron-rich $^{32}$Mg nucleus, which 
has $N=20$, lies at as small as 0.886 MeV, suggesting a deformed shape of 
this nucleus\cite{DGH79,WC80,GMDL84}. 
These experimental data stimulated lots of theoretical 
studies\cite{CFKK75,WBB90}, in which 
the nuclei around $^{32}$Mg have been referred to as the nuclei 
in the {\it island of inversion}\cite{WBB90}. 
That is, the relative position between 
the deformed 2-particle-2-hole intruder state and the spherical 
0-particle-0-hole state is inverted in this region. 
In 1995, Motobayashi {\it et al.} carried out the Coulomb excitation 
measurement of $^{32}$Mg to the first 2$^+$ state, and extracted 
a large $B$(E2) value \cite{MII95}. The large $B$(E2) value is consistent with 
the nuclear deformation (see Fig. 19) \cite{FOS92,UOMH99,UOGMH04}, 
and it has been concluded that the conventional 
$N=20$ magic number does not hold in neutron-rich nuclei (see also 
Refs. \citen{DES06} and \citen{DSA09}). 
A similar disappearance of shell closure has been observed also for 
$N=8$ ~\cite{IMA00} and $N=28$\cite{BGS07}. 

\begin{figure}[t]
\centerline{\psfig{file=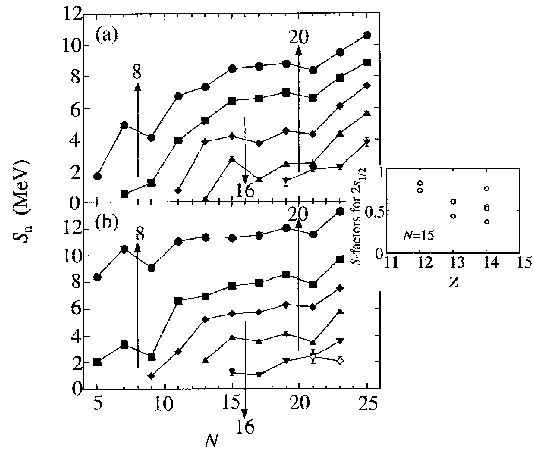,width=7cm}}
\caption{
Systematics of one neutron separation energy, $S_n$, 
for neutron-rich nuclei. 
The upper (a)  and the lower (b) panels correspond to even-$Z$ and odd-$Z$ 
nuclei, respectively. 
The data for nuclei with same $N-Z$ are connected with the lines. 
For the upper (lower) panel, with decreasing order, the lines correspond to 
$N-Z=1,3,5,7,$ and 9 ($N-Z=0,2,4,6,$ and 8). 
A small panel on r.h.s. shows spectroscopic factors of 2s$_{1/2}$ orbit in 
N=15 isotones.
Taken from Ref. \citen{OKSYT00}. }
\end{figure}

A striking finding was that $N=16$ becomes a spherical magic 
number in neutron-rich nuclei, accompanied by the disappearance of 
the $N=20$ magicity. Ozawa {\it et al.} pointed out this fact 
by investigating systematically the neutron separation energy 
in the neutron-rich $p$-$sd$ and the $sd$ shell regions\cite{OKSYT00} 
(see also Refs. \citen{KTO02,CTB02,GBM07,HBB08}). 
When the neutron separation energy 
is plotted as a function of neutron number, it 
suddenly decreases across the shell closure.   
Ozawa {\it et al.} showed that it indeed happens at 
$N=16$ for nuclei with $N-Z>6$  (see Fig. 20) \cite{OKSYT00}, indicating that 
the spherical magic number is shifted from $N=20$ to 16 in neutron-rich 
nuclei. 
Notice that the new magic number $N=16$ implies that the neutron-rich 
$^{24}$O nucleus with $Z=8$ and $N=16$ is a double magic 
nucleus\cite{HBB09,KNP09}. 

It is an important question to ask what makes
the spherical magic numbers change in neutron-rich nuclei. 
Single-particle energies in a Woods-Saxon potential 
already show a quasi-degeneracy 
of 2$s_{1/2}$ and 1$d_{5/2}$ states and that the energy gap at $N=16$ develops 
in weakly-bound systems \cite{OKSYT00,HLZ01}. 
More microscopically, 
Otsuka {\it et al.} have argued that the tensor force as well as 
the spin-spin force of $(\sigma\cdot\sigma)(\tau\cdot\tau)$ type have a 
responsibility for the change of shell structure in neutron-rich 
nuclei\cite{OSF05,OFU01}. 
For instance, the attractive 
spin dependent interaction between the proton 
$d_{5/2}$ orbit and the neutron $d_{3/2}$ orbits leads to a down-shift of the 
neutron $d_{3/2}$ state in stable nuclei such 
as $^{30}_{14}$Si$_{16}$, making the conventional shell gap at $N=20$. 
In the $^{24}_8$O$_{16}$ nucleus, on the other hand, the protons 
do not occupy the $d_{5/2}$ orbit, and thus this attraction does not 
work in the neutron orbits. Consequently, the energy shift of 
the neutron $d_{3/2}$ state does not happen, and the energy gap appears 
at $N=16$ \cite{OSF05,OFU01}. A similar effect may affect also 
the shell structure in heavier regions, {\it e.g.}, the magic numbers 
for superheavy elements \cite{OSU08}. 

\subsection{Deformed halo nuclei} 

 In the vicinity of the stability line, it has been well known 
that many open-shell nuclei are deformed in the
ground state. The nuclear deformation generates the collective
rotational motion, which is characterized by
a pronounced rotational spectrum as well as
strongly enhanced
quadrupole transition probabilities.
As we mentioned in the previous subsection, 
the first evidence for nuclear deformation in neutron-rich 
nuclei was the observation of 
a low-lying state in the $^{32}$Mg nucleus. 
Another well known example is the so called {\it parity inversion} 
phenomenon in $^{11}$Be\cite{TU60}. 
If one used a 
a naive spherical shell model, 
one would expect that the valence neutron in $^{11}$Be occupies the 
1$p_{1/2}$ state in the ground state and the first excited states 
can be constructed by promoting the valence 
neutron to the $sd$ shell, that is, the first excited state would be 
a positive parity state. 
However, the observed ground state of the $^{11}$Be is 1/2$^+$ state 
and the first excited state is 1/2$^-$ state at 0.324 MeV \cite{AB80}. 
This appears as if the 2$s_{1/2}$ and the 1$p_{1/2}$ states are inverted 
in energy. 
This parity inversion problem has been naturally explained by 
considering that $^{11}$Be is a deformed nucleus\cite{EBS95,NTJ96,Hamamoto07}. 

\begin{figure}[t]
\centerline{\psfig{file=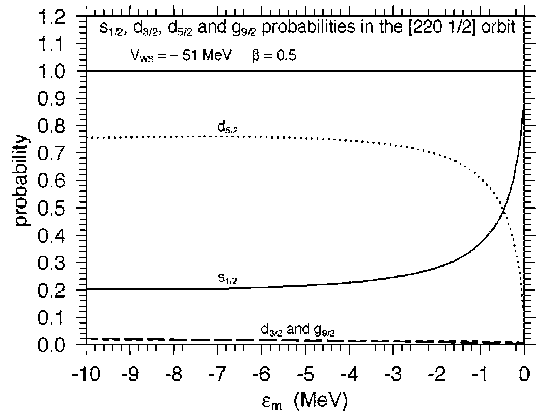,width=7cm}}
\caption{
The probability of each angular momentum component 
in a single-particle wave function in a deformed mean-field potential. 
It is plotted as a function of the corresponding single-particle energy, 
$\epsilon_m$. 
Taken from Ref. \citen{H04}. }
\end{figure}

A single-particle motion in a deformed mean-field potential is 
well known as a Nilsson orbit\cite{BM69,RS80}. 
As a deformed mean-field potential does not have rotational symmetry, 
the corresponding 
single-particle wave functions are obtained as a linear combination 
of several angular momentum components. 
Misu {\it et al.} have pointed out that 
the $s$-wave component, when it contributes, 
becomes dominant in a deformed wave function 
as the separation energy decreases, and eventually 
it gives a 100\% contribution in the limit of zero separation (see Fig. 21) 
\cite{MNA97}. 
They have also shown that the $p$-wave component becomes dominant for negative 
parity states, although it does not give a 100\% contribution even in 
the zero binding limit\cite{MNA97}. See also Refs. \citen{H04} and \citen{YH05} 
for related publications, and Refs. 
\cite{ZMR10,LMR12} for self-consistent mean-field calculations for 
deformed halo nuclei. 

As we discussed in Sec. 2.1, 
the halo structure has been attributed to an occupation of 
$s$ or $p$ single-particle orbit. 
The $s$ and $p$ wave dominance phenomenon in a deformed single-particle state 
suggests that the nuclear deformation 
enhances a chance for a halo formation in weakly bound nuclei. 
The first evidence for a deformed halo was observed recently for the $^{31}$Ne 
nucleus \cite{N09,Takechi12}. 
For this nucleus, both the Coulomb breakup cross section and the interaction 
cross section are found to be large, which clearly suggest a halo structure of 
this nucleus. Although a naive shell model indicates an occupation of the 
1$f_{7/2}$ state by the valence neutron, the halo structure can be 
easily understood by considering a deformation, by which the $p$-wave 
component can be largely coupled in the wave function\cite{H10,UHS11,Minomo}. 
For instance, Urata {\it et al.} have carried out a particle-rotor model 
calculation \cite{EBS95} for this nucleus and shown that 
$I^{\pi}=3/2^-$ is a good candidate for the ground state of $^{31}$Ne, in which 
the component of the $p_{3/2}$ state coupled to the ground state of 
$^{30}$Ne is as large as 44.9\%\cite{UHS11}. 

\subsection{Collective excitations of neutron-rich nuclei}

It has been well known that there are variety of collective 
excitations in atomic nuclei 
\cite{BM69,RS80,BB94,HW01}. 
For stable nuclei, these collective motions are approximately 
classified either as isovector or isoscalar types, in which 
the proton and the neutron motions are out-of-phase or in-phase, 
respectively. 
In neutron-rich nuclei, on the other hand, the 
isovector and the isoscalar modes are coupled to each other due to 
a large asymmetry in proton and neutron numbers \cite{HSZ97,S01}. 
That is, a collective state may have both the isoscalar and isovector 
characters. In the extreme case, 
a pure neutron mode, in which only neutrons contribute to the 
collective excitation, may arise in neutron-rich nuclei\cite{YOF95,M01}. 
A candidate for such neutron mode has been experimentally observed 
in $^{16}$C \cite{WFM08,OIS08,EAD08}. 

There have also been lots of theoretical developments in 
descriptions of collective excitations in neutron-rich nuclei. 
Theoretically, the random phase approximation (RPA) has 
provided a convenient and useful method to describe excited 
states of many-fermion systems\cite{BB94}.
In this method, excited phonon states are described as a superposition of 
many 1-particle 1-hole states. 
For weakly bound nuclei, the continuum effects play an essential role 
due to a much lower threshold energy compared to stable nuclei. 
The continuum RPA method was 
first developed by Shlomo and Bertsch \cite{SB75}, which was subsequently 
applied 
to self-consistent calculations of nuclear giant resonances
with Skyrme interaction by Liu and Van Giai \cite{LG76}.
Hamamoto, Sagawa, and Zhang have extensively applied this method to 
nuclear responses in neutron-rich nuclei\cite{HSZ97,S01,HS96,HSZ97b}. 
An extension of the continuum RPA to deformed nuclei has also been 
carried out by Nakatsukasa and Yabana\cite{NY05}. 
Another important development was to include the pairing effects in 
the continuum RPA. As we discussed in Sec. 3, the pairing and the continuum 
couplings play an essential role in neutron-rich nuclei. 
These effects can be taken into account by extending the RPA to the 
quasi-particle RPA (QRPA). The continuum QRPA 
method on top of the ground state described by 
the Hartree-Fock-Bogoliubov method has been 
developed by several 
groups\cite{M01,KSGG02,KSGG04,PRNV03,YG04,TEBDNS05,YYM06,
PG08,YG08,LPD10,DR11,SM11}, 
and has been applied to neutron-rich nuclei. 

\begin{figure}[t]
\centerline{\psfig{file=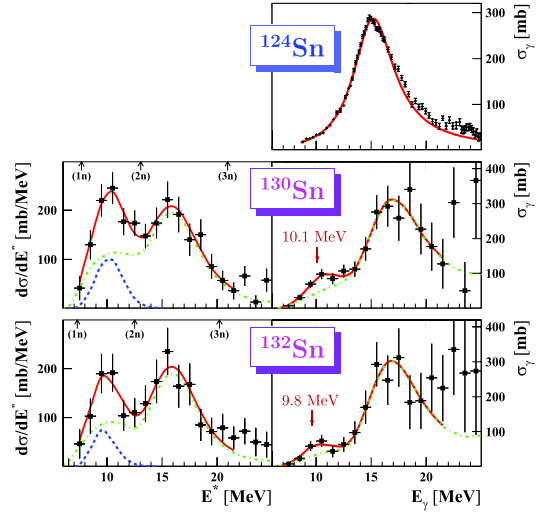,width=7cm}}
\caption{
The experimental data for the Coulomb excitation (the left panels) 
and the photo-neutron (the right panels) cross sections 
for $^{124}$Sn, $^{130}$Sn, and $^{132}$Sn. 
Taken from Ref. \citen{AKF05}.}
\end{figure}

Recently, much attention has been paid to low-lying dipole (E1) strength in 
neutron-rich nuclei, which has often been referred to as 
{\it pygmy dipole resonance}.  
It has been well known that the dipole strength is far dominated by 
the giant dipole response (GDR)\cite{BB94,HW01} (see also 
Ref. \citen{Tamii11} for a recent complete measurement for the dipole strengths 
in $^{208}$Pb over a wide range of excitation energy). 
However, nuclei with a neutron excess often show a low-lying dipole strength 
at energies much lower than the GDR. 
The pygmy dipole resonances have been experimentally found not only in 
neutron-rich nuclei\cite{LAB01,TAB02, TBH03,GBM08,WBC09,AKF05,KPA07} 
but also in stable nuclei 
\cite{HEMVVZ04,GBB98,RHK02,EBE03,HBE97,HFB99,ZVB02,VTB06
,SBB08,SFH08,THK10} (see Fig. 22). 
The pygmy dipole resonance is important also from the astrophysical point 
of view, as it affects significantly a radiative neutron capture rate, 
which is relevant to the r-process nucleosynthesis\cite{GKS04}. 
Although the exact nature of the pygmy dipole resonance has not yet 
been 
clarified completely, it has been pointed out that the pygmy dipole strength 
is strongly correlated with the neutron skin 
thickness\cite{P06,INY11} (see, however, also a counter argument in 
Ref. \citen{RN10}). As the neutron skin thickness is intimately 
related to the equation of state (EOS) in asymmetric nuclear 
matter\cite{YS04}, it is expected that the information on the 
nuclear matter properties, such as the symmetry energy coefficients, 
may be obtained by studying the pygmy dipole strength in neutron-rich 
nuclei\cite{KPA07}. 

\section{Summary}

Physics of unstable nuclei has been developed rapidly thanks to the 
recent availability of radioactive beams in the world. 
A new era has been commenced in nuclear physics, where the 
isospin degree of freedom, that is one of the fundamental 
quantum numbers in atomic nuclei, can be controlled in self-bound 
interacting Fermion systems. 
Many new features of atomic nuclei have been discovered or theoretically 
discussed so far, 
as we have reviewed in this Chapter. 
We list some of these below: 

\begin{enumerate}

\item 
The most prominent discovery was the halo structure, in which the 
neutron density distribution largely extends over the proton distribution. 
In stable nuclei, the neutron and the proton distributions are similar 
to each other within a scaling factor, and thus the halo structure 
is a clear manifestation of the decoupling of proton and neutron in 
weakly bound nuclei. The halo structure is ascribed to an occupation of 
either $s$ or $p$ orbits, for which the root mean square radius diverges 
in the zero binding limit. Typical examples of the halo nuclei include 
$^6$He, $^{11}$Li, and $^{11}$Be. $^{31}$Ne has also been identified recently 
as a deformed halo nucleus. 

\item
A similar decoupling effect is the skin structure. While the halo structure 
corresponds to a long low-density tail of neutron distribution, the skin 
structure corresponds to a layer of extremely neutron-rich matter. 
The matter radii of neutron-rich nuclei have been systematically 
studied with interaction cross section and isotope shift measurements. 

\item 
Neutron-rich nuclei often show a soft dipole excitation, that is, 
dipole strengths in the low excitation energy 
region. For halo nuclei, the soft dipole mode is attributed to the 
threshold effect, that is, the optimal matching of wave functions 
between a weakly bound and continuum states. For skin nuclei, the low-lying 
dipole resonances have been referred to as the pygmy dipole mode. 
Even though the exact nature of the pygmy dipole mode has not yet been 
fully clarified, it has been expected that it is closely related 
to the equation of state of asymmetric nuclear matter. 

\item 
The dineutron correlation in neutron-rich nuclei, 
such as $^{11}$Li and $^6$He,  
has been theoretically predicted. 
This is a spatial correlation, with 
which two valence neutrons take a compact configuration. The recent 
experimental data for the Coulomb dissociation of $^{11}$Li strongly 
suggests the existence of the dineutron correlation in $^{11}$Li. 
A similar correlation in proton rich nuclei, that is, the diproton 
correlation has also been predicted. The two-nucleon radioactivities 
as well as the two-neutron transfer reactions are expected to provide 
a direct probe of the dineutron correlation. 

\item 
Another important feature in neutron-rich nuclei is the change of 
(spherical) magic numbers. It was considered that 
the magic numbers exist independently for proton and neutron. 
However in neutron-rich nuclei some of the magic numbers have been 
found to disappear ($N$=8 and 20) and a new spherical 
magic number appears ($N$=16). These changes have revealed the importance 
of tensor interaction. 

\end{enumerate}

A new generation RI beam facility, RIBF, at RIKEN, Japan, has already 
been in operation, 
and other new generation facilities, such as 
FAIR (Germany), SPIRAL2 (GANIL), and FRIB (USA), will also be in operation 
in a few years. 
We are now at a stage in which 
we gain a deep insight in nuclear many-body systems, from 
stable nuclei to weakly-bound unstable nuclei in a unified manner, 
and are about to 
explore a vast {\it terra incognita} \cite{NC01} in the nuclear chart.

\section*{Acknowledgments}

We thank A. Vitturi, P. Schuck, H. Esbensen, G. Colo, J. Margueron, 
T. Oishi, Y. Urata, 
T. Nakamura, S. Shimoura, H. Sakurai, 
A. Navin, M. Takechi, and L. Corradi for collaborations and many 
useful discussions. 
This work was supported 
by the Japanese
Ministry of Education, Culture, Sports, Science and Technology
by Grant-in-Aid for Scientific Research under
program no. (C) 22540262.

\end{document}